\pdfoutput=1

\documentclass[usenatbib]{mn2e}

\usepackage[utf8]{inputenc}  
\usepackage[T1]{fontenc}     
\usepackage[english]{babel}  

\usepackage{amsmath}
\usepackage{xspace}     
\usepackage[section,below]{placeins}   
\usepackage{subfigure}  
\usepackage{hyperref}

\usepackage{graphicx}
\DeclareGraphicsExtensions{%
  .pdf,.PDF,%
  .png,.PNG,%
  .jpg,.mps,.jpeg,.jbig2,.jb2,.JPG,.JPEG,.JBIG2,.JB2 %
}

\newcommand{\spl}{SpaghettiLens\xspace}
\newcommand{\sw}{Space Warps\xspace}

\newcommand{\ERf}[1][]{$\Theta_\text{E#1}$\xspace} 

\newcommand{\figref}[1]{Figure~\ref{fig:#1}}
\newcommand{\secref}[1]{Section~\ref{sec:#1}}
\newcommand{\tabref}[1]{Table~\ref{tab:#1}}
\newcommand{\Figref}[1]{Figure~\ref{fig:#1}}

\newcommand{\tgeom}{t_{\rm geom}}
\newcommand{\tgrav}{t_{\rm grav}}
\newcommand{\subcirc}{{\lower.33ex\hbox{$\circ$}}}
\newcommand{\subbullet}{{\lower.33ex\hbox{$\bullet$}}}

\usepackage{color}

\newlength{\myplotswidth}
\begin{document}

\title{Gravitational Lens Modelling in a Citizen Science Context}

\date{Accepted: 2014 Nov. 30. Received 2014 Nov. 05. In original form 2014 Dec. 02.}

\pagerange{2170--2180}
\pubyear{2015}
\volume{447}

\author[Küng et al]{Rafael Küng,$^{1}$
Prasenjit Saha,$^{1}$
Anupreeta More,$^{2}$
Elisabeth Baeten,$^{3}$
\newauthor
Jonathan Coles,$^{4}$
Claude Cornen,$^{3}$
Christine Macmillan,$^{3}$
Phil Marshall,$^{5}$ 
\newauthor
Surhud More,$^{2}$
Jonas Odermatt,$^{6}$
Aprajita Verma$^{7}$
and Julianne K. Wilcox$^{3}$
\\
$^{1}$Physik-Institut, University of Zurich, Winterthurerstrasse 190, 8057 Zurich, Switzerland\\
$^{2}$Kavli Institute for the Physics and Mathematics of the Universe, University of Tokyo, 5-1-5 Kashiwanoha, Kashiwa-shi 277-8583, Japan\\
$^{3}$Zooniverse, c/o Astrophysics Department, University of Oxford, Oxford OX1 3RH, UK \\
$^{4}$Exascale Research Computing Lab, Campus Teratec, 2 Rue de la Piquetterie, 91680 Bruyeres-le-Chatel, France\\
$^{5}$Kavli Institute for Particle Astrophysics and Cosmology, Stanford University, 452 Lomita Mall, Stanford, CA 94035, USA\\
$^{6}$Kantonsschule Zug, L\"ussiweg 24, 6300 Zug, Switzerland\\
$^{7}$Sub-department of Astrophysics, University of Oxford, Denys Wilkinson Building, Keble Road, Oxford, OX1 3RH, UK\\
}

\maketitle

\begin{abstract}

We develop a method to enable collaborative modelling of
gravitational lenses and lens candidates, that could be used by
non-professional lens enthusiasts.  It uses an existing free-form
modelling program (GLASS), but enables the input to this code to be
provided in a novel way, via a user-generated diagram that is
essentially a sketch of an arrival-time surface.

We report on an implementation of this method, \spl, that has been
tested in a modelling challenge using 29 simulated lenses drawn from a
larger set created for the \sw citizen science strong lens search. We
find that volunteers from this online community asserted the image
parities and time ordering consistently in some lenses, but made
errors in other lenses depending on the image morphology. While errors
in image parity and time ordering lead to large errors in the mass
distribution, the enclosed mass was found to be more robust: the
model-derived Einstein radii found by the volunteers were consistent
with those produced by one of the professional team, suggesting that
given the appropriate tools, gravitational lens modelling is a data
analysis activity that can be crowd-sourced to good effect. Ideas for
improvement are discussed; these include (a)~overcoming the tendency
of the models to be shallower than the correct answer in test cases,
leading to systematic over-estimation of the Einstein radius by 10\%
at present, and (b)~detailed modelling of arcs.

\end{abstract}

\begin{keywords}
gravitational lensing: strong, methods: numerical
\end{keywords}


\section{Introduction}

The first work on lens modelling
\citep{1981ApJ...244..723Y,1981ApJ...244..736Y} was developed after
the discovery of the first two gravitational lenses
\citep{1979Natur.279..381W,1980Natur.285..641W}, where a massive
galaxy causes a background quasar to appear as two or four
images.  For the second lens to be discovered (PG1115+080), mass
models scored an early success with the prediction that one of the
lensed images seen would split further into two at higher resolution.
That galaxies must sometimes cause multiple images had long been
expected \citep{1937ApJ....86..217Z}, and it had even been argued that
the phenomenon could help measure cosmological parameters
\citep{1964MNRAS.128..307R,1966MNRAS.132..101R}, but apparently nobody
was expecting that lenses would need detailed modelling.  The first
observations, however, immediately stimulated models.  The reason for
that lies in the image separation. Recall that image separations are
of order of the angular Einstein radius
\begin{equation}
\Theta_{\rm E}
\sim \left(\frac{4GM}{c^2 D_L}\right)^{1/2}
\simeq 0.1'' \left(\frac{M}{M_\odot}\right)^{1/2}
             \left(\frac{D_L}{\rm pc}\right)^{-1/2},
\end{equation}
where $D_L$ is the distance to the lens, and $M$ its mass.  A lensing
galaxy with $M\sim10^{11}\rm\,M_\odot$ at $\sim1\rm\,Gpc$ would cause image
separations of $\sim1''$, which is comparable to the size of the
galaxy; typically the lensed images are seen through the galaxy halo.
Hence, the lensed images depend on the detailed mass distribution of
the lensing galaxy.  Galaxy lenses therefore require models of their
mass distributions.

Since those early discoveries, more than 400 secure lenses are now
known. Modelling of the mass distribution is part of any research
using lenses, but so far no modelling study has spanned all known
lenses.  The largest single one \citep{2009ApJ...703L..51K} models 58
separate lenses to infer the distribution of dark matter around
galaxies.  In other work, \cite{2011ApJ...740...97L} combined lens
models of 21 galaxies with models of their stellar populations, to
find the relation between stars and dark matter, and
\cite{2014MNRAS.437..600S} modelled 18 time-delay lenses together to
infer cosmological parameters.

Imaging surveys now under way aim to increase the inventory of lenses
another ten or a hundred fold \citep[see
e.g.][]{Marshall2005,OguriMarshall2010}, with both automated and
visual search techniques proposed
\citep[e.g.][]{Marshall2009,More2012ApJ,Gavazzi2014}. For example, \sw
(Marshall et al, in prep; More et al, in prep) is a citizen science
project\footnote{\url{http://www.spacewarps.org}} in which volunteers
are presented sky-survey images and are invited to identify lens
candidates, by eye.  Simulated lenses are mixed in with the data,
both to help train volunteers on what to look for, and to provide
measures of the effectiveness of the search.  The motivation for \sw
is to enable volunteers, some of whom had previously serendipitously
identified lens candidates on earlier citizen-science surveys,
either to make discoveries missed in automatic searches by
software robots, or to perform the necessary inspection of an
automatically-generated sample, for quality control.  Robots can be
built to be good at detecting lensing system in clean, uncrowded fields
with high signal-to-noise, but in more general test situations, robots
miss lenses (low completeness) or contaminate the results with
non-lenses (low purity) \citep{Marshall2009}.

The encouraging early results from the first \sw lens search, carried
out on the $\simeq172$ square degree Canada-France-Hawaii Telescope
Legacy Survey (CFHTLS) imaging by over 30,000 volunteers (Marshall et
al, in prep; More et al, in prep) prompted the question: could the
modelling of the lenses also be done by the volunteers?  If so,
modelling could help prioritise lens candidates at an early stage,
which would be very useful with new wide-field and sensitive surveys,
which will yield thousands of lens candidates.  There are several
software tools for lens modelling available, and work has been done on
generic interfaces \citep[e.g.][]{2014A&C.....5...28L}.  Some early
designs for \sw included a prototype lens modelling tool
\citep{2010AAS...21543527N}. Moreover, some \sw volunteers are quite
experienced from earlier projects, having individually spent a
thousand hours or more with data, and are very interested in more
demanding projects.  The interests of citizen-science communities are
just beginning to be studied \citep[e.g.,][]{2013AEdRv..12a0106J}, but
it is clear that some volunteers welcome open-ended challenges, and
sometimes these have led to new scientific results: one example is the
discovery of an exceptional extra-solar planet
\citep{2013ApJ...768..127S}; another is the development of new
algorithms for protein folding \citep{Khatib22112011}.  All these are
grounds for optimism.  There is, however, a basic difficulty in strong
gravitational lensing. Lensed images do not look much like their
source, and still less do they resemble the lensing-mass distribution.
To model a lens, one needs either to do a lot of random guessing, or
to have a good intuition for what works.

In this paper, we propose a way around the difficulty, and report on a
modelling test on \sw using simulated lenses.  The three following
sections are devoted to the concept, the implementation, and tests
respectively.

In \secref{Fermat} we introduce a markup system for lensed images,
which we call a ``spaghetti diagram.''  A spaghetti diagram resembles
the visible image system, in a cartoon-like way, and at the same time
it encodes the basis of a mass model.  This supplies an intuitive link
between the image system and the mass distribution, which look
frustratingly different from each other.  Spaghetti diagrams are
essentially the saddle-point contours originally introduced to
gravitational lensing by \cite{1986ApJ...310..568B} as a way of
classifying lensed images.  They are sometimes shown as part of the
output of lens models \citep[for
example][]{2001ApJ...557..594R,2003ApJ...590...39K,Lubini2012}.  In
the present work, however, spaghetti diagrams are the {\em input\/}
through which the modeller tells the \spl program what to do.

In \secref{SpaghettiLens} we describe the \spl program, which
implements the above scheme.  \spl is an interface to and
extension of the GLASS framework for modelling lenses
\citep{2014arXiv1401.7990C}.  We will not go into software details in
this paper, instead concentrate on lens modelling per se, but we
remark that \spl is designed to be friendly to the forum style of
citizen-science projects, and enables incremental collaborative
model refinement by different people, without sacrificing any of the
technical features of GLASS.

In \secref{mod_challenge} we describe a modelling challenge where a
diverse sample of 29 simulated lenses was modelled multiple times
by a small number of \sw volunteers using \spl.  The models
were then examined in two ways.  One was whether the spaghetti diagram
was correct.  The other was the recovery of the Einstein radius of the
lens.  In addition, we show some visual comparisons of the actual and
recovered lens shape and radial profile, and identify some areas to
improve. Profile and shape recovery with GLASS has been studied in
more detail in \citep{2014arXiv1401.7990C}.

\secref{todo} gives the general outlook and next steps.


\section{Fermat's principle and spaghetti diagrams} \label{sec:Fermat}

We first explain the lensing theory relevant to \spl, following the
formulation of gravitational lensing in terms of Fermat's principle by
\cite{1986ApJ...310..568B}.


\subsection{Geometrical and gravitational time delays}

Consider a lens at some redshift $z_L$ and let $(x,y)$ be planar
coordinates at the lens, transverse to the line of sight.  Let
$\Sigma(x,y)$ be the mass distribution.  It is a mass per unit area,
i.e., density projected along the line of sight.  The mass
distribution is often given in a dimensionless form
\begin{equation} \label{eq:kappa}
\kappa(x,y) \propto \Sigma(x,y)
\end{equation}
called the convergence.  Let there be light, in the form of a more
distant source, at redshift $z_S$, behind point $(x_s,y_s)$ on the
lens.

We now imagine a virtual photon flying from the source to some $(x,y)$
on the lens, then changing direction and coming to the observer.  Such
a direction change would increase the light travel time compared to
coming through $(x_s,y_s)$.  The increased light travel time from the
geometry of deflection would be
\begin{equation} \label{eq:tgeom}
\tgeom(x,y) \propto (x-x_s)^2 + (y-y_s)^2 ,
\end{equation}
assuming the delay is small compared to the total light travel time.

An additional delay of the photon comes from travelling through the
curved spacetime at the lens.  This gravitational time delay $\tgrav$
is related to the mass distribution of the lens.  The relation is
generally written as a two-dimensional Poisson equation, but an
alternative expression, avoiding calculus, is as follows.  The value
of $\tgrav$ through $(x,y)$ equals its average value on the
circumference of a small circle centred at $(x,y)$, plus a constant
times the mass within that circle.  The constant is $2G/c^3$ times the
cosmological expansion factor $(1+z_L)$. Thus
\begin{equation} \label{eq:tgrav}
\tgrav(x,y) = \left\langle \tgrav(x_\subcirc,y_\subcirc) \right\rangle
              + (1+z_L) \frac{2G}{c^3} M(x_\subbullet,y_\subbullet) \,.
\end{equation}
We have used $(x_\subcirc,y_\subcirc)$ to denote the circumference of
a circle, and $(x_\subbullet,y_\subbullet)$ to indicate the integrated
mass within the circle.  Appendix~\ref{more-theory} relates this
expression to the better-known explicit form for the gravitational
time delay.

The light travel time of a virtual photon is therefore longer by
\begin{equation}  \label{eq:tarriv}
t(x,y) = t_{\rm geom} + t_{\rm grav}
\end{equation}
than it would have been with no lens present.  Real photons take paths
that make $t(x,y)$ extremal, that is, having a minimum, maximum or
saddle point (Fermat's principle).

The proportionality factors in \eqref{eq:kappa} and \eqref{eq:tgeom}
depend on the redshifts and cosmological parameters, and are given in
Appendix~\ref{more-theory}.

\begin{figure}
\centering
\includegraphics[width=0.95\columnwidth]{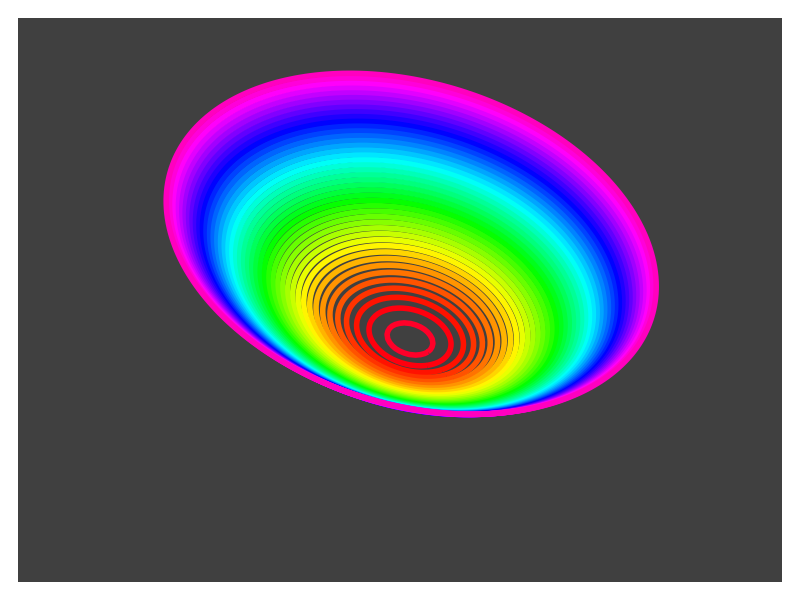}
\includegraphics[width=0.95\columnwidth]{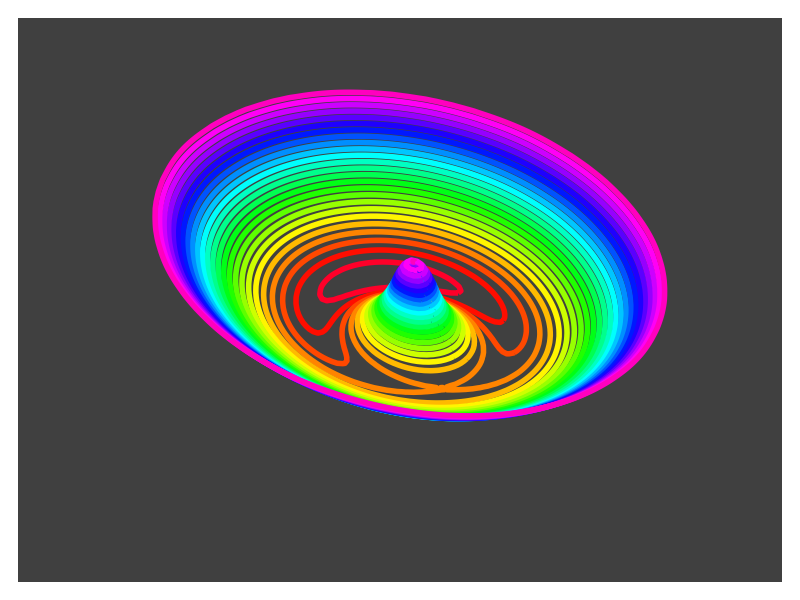}
\includegraphics[width=0.95\columnwidth]{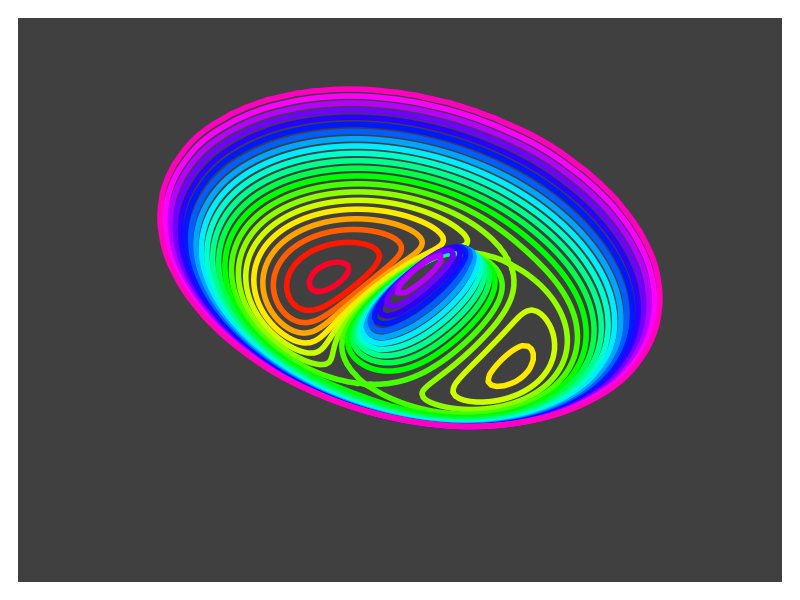}
\caption{Perspective views and contour maps of example arrival time
  surfaces.  Contours are coloured in rainbow order (red: least delay,
  violet: highest delay).  The special contours that self-cross at
  saddle points are the basis of spaphetti diagrams.  {\em Upper
    panel:\/} No lens, hence showing the parabolic shape of the
  geometrical time delay.  The image would be at the bottom,
  coinciding with the source. {\em Middle panel:\/} A circular lensing
  mass (offset from the source) has been added, which has pushed the
  minimum to one side and introduced a maximum and saddle point, each
  corresponding to an image.  The saddle-point is characterized by a
  self-crossing ``spaghetti'' contour. {\em Lower panel:\/} An
  elongated lensing mass has been added.  There are now two minima,
  two saddle points, and a maximum, each corresponding to an image.}
\label{fig:arriv}
\end{figure}


\subsection{Arrival-time contours} \label{sec:arriv}

The full function $t(x,y)$, also known as the arrival-time surface,
applies to virtual photons.  In other words, it is an abstract
construct and not itself observable.  But observable image positions
can be derived from the arrival-time surface, so visualising the
surface is useful.  \figref{arriv} does so.  In this figure, a
maximum, if present, is easy to see.  To locate mimima and
saddle-points, however, one needs to examine the contours of equal
arrival time.  A saddle point is characterised by a contour crossing
itself, forming an~X.  Mimima, on the other hand, have contours
looping around them, as do maxima.

The saddle-point contours which form an X are especially interesting,
because they set the overall topography of the arrival-time surface.
They obviously give the locations of the saddle points, and roughly
localise the minima and maxima as well.  If more precise locations of
the minima and maxima are added, the whole arrival-time surface is
already approximately known.  Since the arrival-time surface has an
exact relation to the lens-mass distribution and the source position,
in effect the mass distribution is also automatically approximately
specified.  In other words, a simple sketch of saddle-point contours
along with locations of minima and maxima ---which we call a
``spaghetti diagram''--- is implicitly already an approximation to a
lens-mass distribution.

The preceding assumes a point source.  To get an idea of what an
extended source would do, let us imagine moving the original source
slightly.  The contours of constant arrival time will naturally move
slightly, and so will the images.  The movement of the contours will
be most noticeable where the contours are far apart, that is where the
arrival-time surface is nearly flat.  As is evident from
\figref{arriv}, this is the region where the minimum and saddle points
lie, or near the images.  In this region, points on the source that
are close together produce images that are comparatively far apart.
In other words, the image is highly magnified.  In summary, lower
curvature in the arrival-time surface for a point source implies
larger magnification of an extended source.  Conversely, where the
arrival-time surface is strongly curved, the image will be
demagnified.  We see from \figref{arriv} that the arrival-time surface
tends to be highly curved near the maximum.  Hence maximum tend to be
demagnified.  In practice, maxima of the arrival time are nearly
always too faint to see. The minima and saddle points dominate.


\section{A lens-modelling program}
\label{sec:SpaghettiLens}

\spl is a mass modelling program that makes use of the \sw
infrastructure, in particular, the image database and the discussion
forum.\footnote{\url{ http://talk.spacewarps.org}} The forum is
essential for establishing contact between interested members of the
\sw community and the project science team, and then for enabling
collaboration between them. We were able to collaborate together on
modelling objects from \sw in the usual style of medium-sized
astronomical collaborations, with video-conferencing and in-person
meetings where possible. Preliminary results were immediately
summarized on modelling threads on the forum, and anyone interested
was made welcome to join at any time.

Modelling with \spl involves three stages, (1)~markup of the image,
followed by (2)~intensive numerical computation carried out on a
server in the background, followed by (3)~review of diagnostics and
possible discussion.  Human interaction is essential to the first and
third stages, while stage~2 is completely automated. We now describe
the three stages.


\subsection{Image markup}

\begin{figure}
  \centering
    \includegraphics[width=0.90\linewidth]{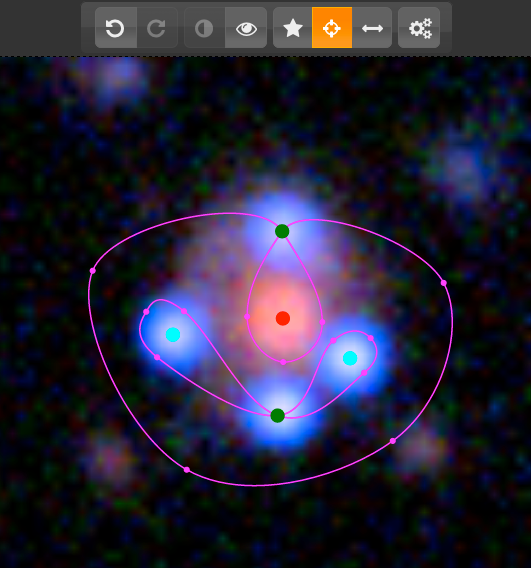}
  \caption{Screen grab of \spl in action. A \sw image has been loaded
    in, re-centred and zoomed.  Five images and the associated
    ``spaghetti'' contours have then been suggested, using the marking
    tools associated with the buttons along the top of the panel. The
    mass model is generated server-side when the right-most button is
    pressed.}
  \label{fig:screenshot}
\end{figure}

One begins by going to the \spl web application\footnote{\url{
http://mite.physik.uzh.ch}} and entering the number of a
\sw image tile.  \spl then presents the image, along with zoom and pan
options and a markup tool to construct a spaghetti diagram.  The human
modeller now has to make an educated guess for the topography of the
arrival-time surface, and input the locations time-ordering of the
maxima, minima, and saddle-points.  The markup tool \citep[which is
inspired by Figure~6 of][and is like that figure made interactive and
overlaid on data]{1986ApJ...310..568B} lets the modeller enter the
information by sketching saddle-point contours.  Examples can be seen in Figure~\ref{fig:screenshot} and the upper-left
panels of Figures \ref{fig:6941} to \ref{fig:7022}.  The loops in the
markup tool were the origin of the ``spaghetti'' metaphor.

The markup tool allows only valid lensing configurations to be
entered.  The user does not need to think explicitly about the image
parities (though the markup tool provides this information using
colour codes) or about time-ordering, or worry about the odd-image
theorem.  The exact placement of the loops in a spaghetti diagram has
no significance.  Only the hierarchy of which loop is inside which is
relevant.  The loops are there simply to help modeller's intuition.

As implemented so far, \spl assumes that the lens is dominated by a
single galaxy.  Accordingly, only one maximum in the arrival-time surface
is permitted, and it is taken to be the centre of the main lensing
galaxy.  The user can, however, mark additional minor galaxies: these
are modelled as point masses, the mass being fitted by the program along
with the rest of the mass distribution.


\subsection{Numerics}

Having sketched a spaghetti diagram, the user presses a button to
initiate the next stage.  \spl then translates the spaghetti diagram
into input for GLASS, and forwards this input.  The task of GLASS,
which runs server-side as it is compute-intensive, is to find a mass
distribution $\kappa(x,y)$ that exactly reproduces the given locations
of the maximum, minima and saddle points. This criterion by itself is
extremely under-determined --- there are infinitely many mass
distributions that will reproduce a given set of maxima, minima and
saddle points, but typically they (a)~produce lots of extra images,
and (b)~look very unlike galaxies.  Additional assumptions (a prior)
are necessary.  GLASS uses the following priors
\citep[cf.][]{1997MNRAS.292..148S,2008ApJ...679...17C}.
\begin{enumerate}
\item The mass distribution is built out of non-negative tiles of
  mass.  (Sometimes these tiles are called mass pixels, but we should
  emphasize that they are unrelated to image pixels, and are much
  larger.)
\item There is a notional lens centre, say $(x_0,y_0)$ which is
  identified with the maximum of the arrival time.  The source can
  have an arbitrary offset with respect to the lens centre.
\item The mass distribution must be centrally concentrated, in two
  respects.  First, the circularly averaged density must fall away
  like $$ \left[(x-x_0)^2+(y-y_0)^2\right]^{-1/2}$$ or more steeply.
  Second, the direction of increasing density at any $(x,y)$ can point
  at most $45^\circ$ away from $(x_0,y_0)$.
\item The lens must be symmetrical with respect to $180^\circ$ rotations
  about $(x_0,y_0)$.  This symmetry assumption can be relaxed if the
  user wishes.
\end{enumerate}
There are still infinitely many models that satisfy both data and
prior constraints, but now they are more credible as galaxy lenses.
It is then possible to generate an ensemble of models.  The sampling
technique used by GLASS is described in \citep{Lubini2012}.
Typically, ensembles of 200 models are used.  That is to say, what we
call a \spl model is really the mean of an ensemble of 200 models, and
its estimated uncertainty is the range covered by the whole ensemble.


\subsection{Diagnostics}
\label{sec:diag}

After the model ensemble has been generated, \spl post-processes it to
present results and diagnostics to the user for inspection. This takes
the form of three figures.
\begin{enumerate}
\item A synthetic image of the lensed features.
\item A contour map of the arrival-time surface $t(x,y)$.
\item A gray scale plus contour map of the mass distribution.
\end{enumerate}
The synthetic image generated by \spl assumes a simple
circularly symmetric source with linearly decreasing
surface brightness profile.  The user can change the contrast level
on the image, which (though it is not saved) amounts to adjusting the
size of the source. These synthetic images are still very
crude, and not always useful for assessing models.  The best
indicator, in practice, of whether the modelling was successful is
contour map of $t(x,y)$, with saddle-point contours highlighted.  It
is, in effect, the computer's refinement of the spaghetti diagram
input by the user.  If the arrival-time surface looks qualitatively
similar to the spaghetti diagram, that generally indicates a
successful model.  The mass distribution also provides indications;
successful models generally lead to smooth-looking mass distribution,
whereas an irregular or checkboard pattern in the mass map signals a
bad model.

After examining this feedback, the user can choose to save the
model to the \spl archive, at which point it is assigned an unique
URL.  They can also modify the input and try again, or discard the
attempt altogether.  After archiving, there can be discussion among
modellers, through the \sw forum or by any other means, and revision
of the model.  This is achieved simply by sharing the model's
URL; following its hyperlink takes one to the \spl app, pre-loaded
with the correct data image and input spaghetti. Any archived model
can be revised by any user: they can modify the spaghetti
configuration slightly or drastically, or change options like the size
of the mass tiles. Particularly interesting lens candidates lead to
trees of models in this way.  Discussion among modellers tends to
prune a model tree, focusing attention on the most interesting
models.\footnote{See ``Collaborative gravitational lens
modelling\dots'' in {\tt http://letters.zooniverse.org} for an example.}


\section{A lens modelling challenge} \label{sec:mod_challenge}

We now describe a test of the lens-modelling system, under conditions
that mimic as closely as possible the modelling of real lens
discoveries.  The lenses to be modelled were the simulated lenses
(known as ``sims'') already sprinkled onto the \sw field.  Once a
small user base had grown around \spl, a modelling challenge was
announced through the \sw forum.  The challenge set consisted of 29
sims, chosen to represent the different visual morphologies of \sw
sims. Modellers then contributed 119 models for these sims (at least
two for each sim).  Models were reported on the same forum used to
model real candidate lenses.  Modellers were free to consult and
refine each other's models, but had no information on how the sims
were generated.

Once the modelling was complete, the models were compared with the
originals.  There were two main tests: a check of whether the
spaghetti diagrams were correct for the lens in question, and a
comparison of the effective Einstein radii of the sims and the models.


\subsection{The simulated lenses} \label{sec:sims}

The \sw sims are described in detail in More et al (in prep), but
relevant here is that the sims were of three kinds, as follows.
\begin{enumerate}
\item Lensed quasars: The lens is modelled as a singular isothermal
  ellipsoid (SIE) and a constant external shear whereas the quasar is
  represented with a circular Gaussian source whose size is given by
  the point spread function (PSF) in each imaging band.
\item Galaxy-scale lenses: The lens model is the same as above whereas
  the background galaxy is modelled as an elliptical de~Vaucouleurs.
\item Group-scale lenses: The lens model includes SIE models for the
  central galaxy and the inner group members, plus a circular NFW
  \citep{1996ApJ...462..563N,1997ApJ...490..493N} to represent the
  underlying dark matter distribution and the background galaxy model
  stays the same as galaxy-scale lenses.
\end{enumerate}

\FloatBarrier

\begin{figure}
  \centering
  \includegraphics[width=\myplotswidth]{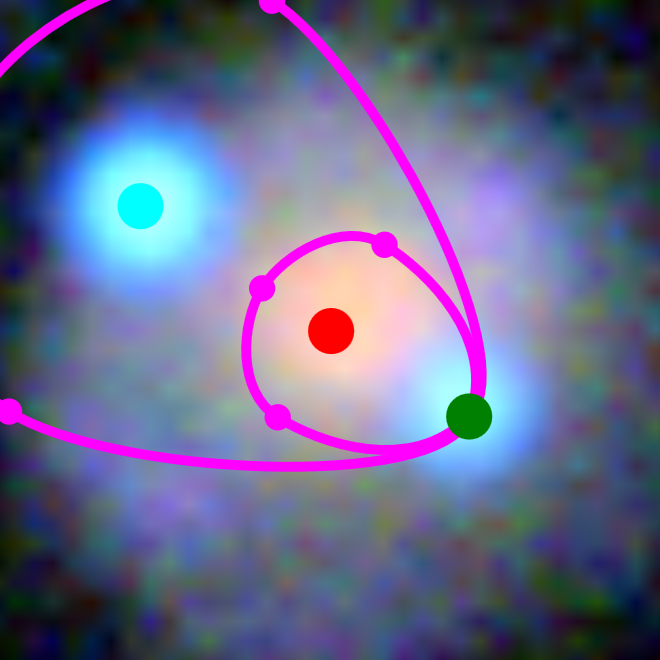}
  \includegraphics[width=\myplotswidth]{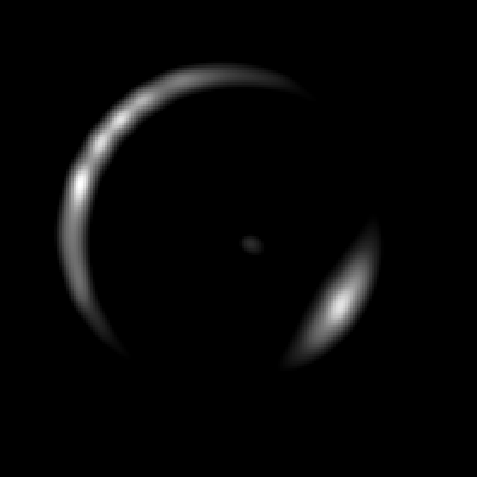} \\
  \includegraphics[width=\myplotswidth]{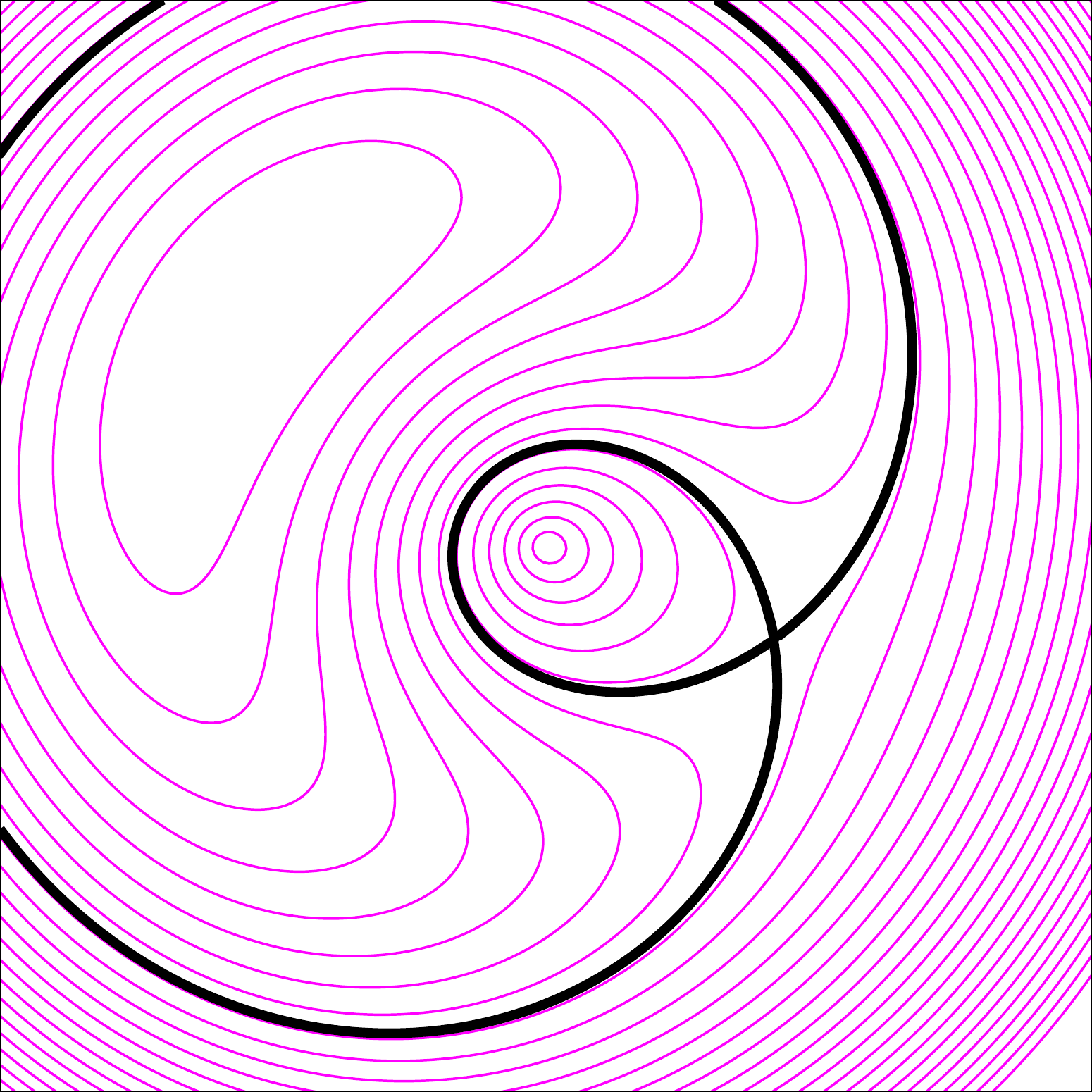}
  \includegraphics[width=\myplotswidth]{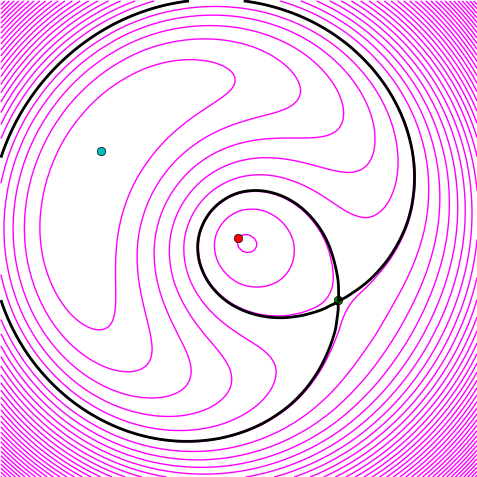} \\
  \includegraphics[width=\myplotswidth]{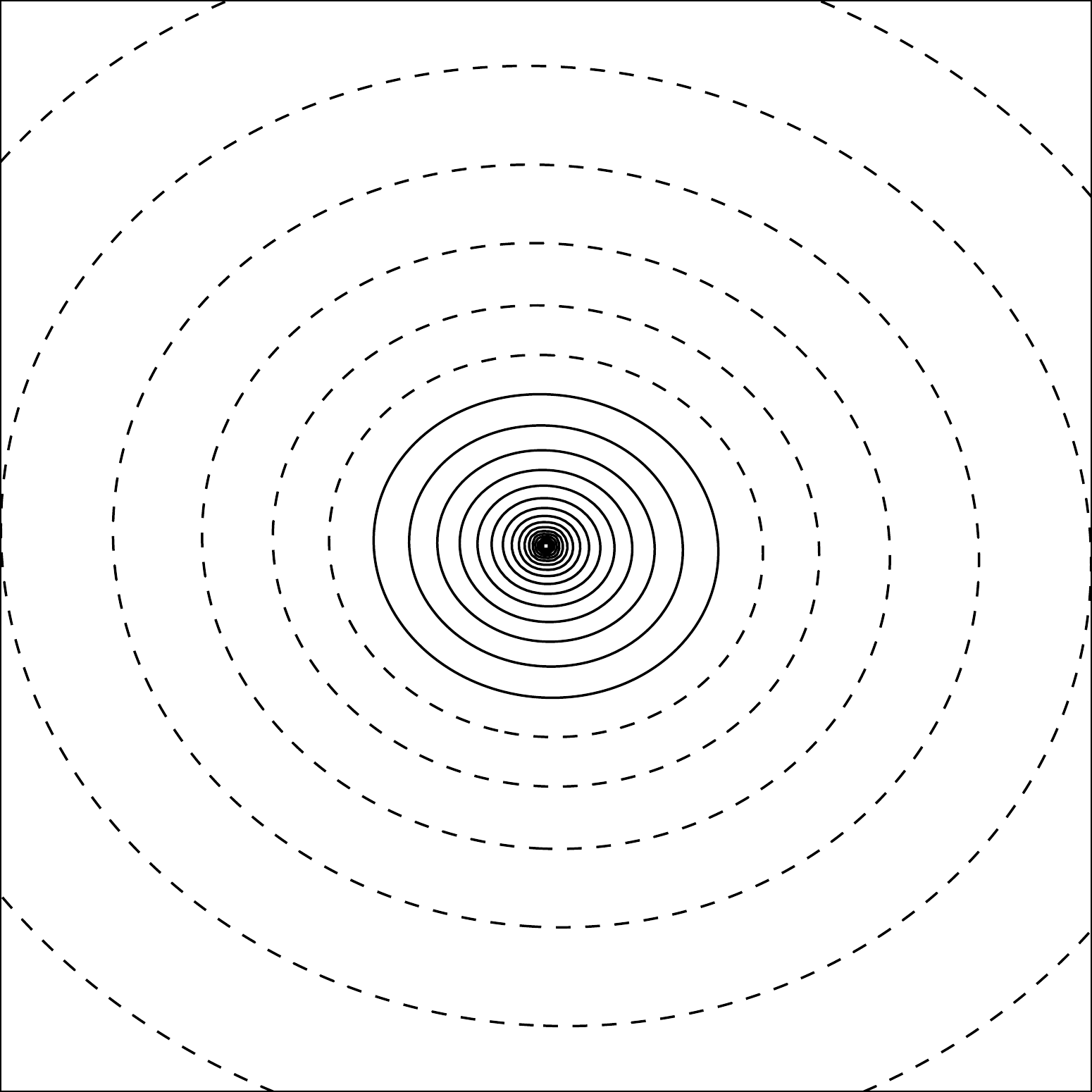}
  \includegraphics[width=\myplotswidth]{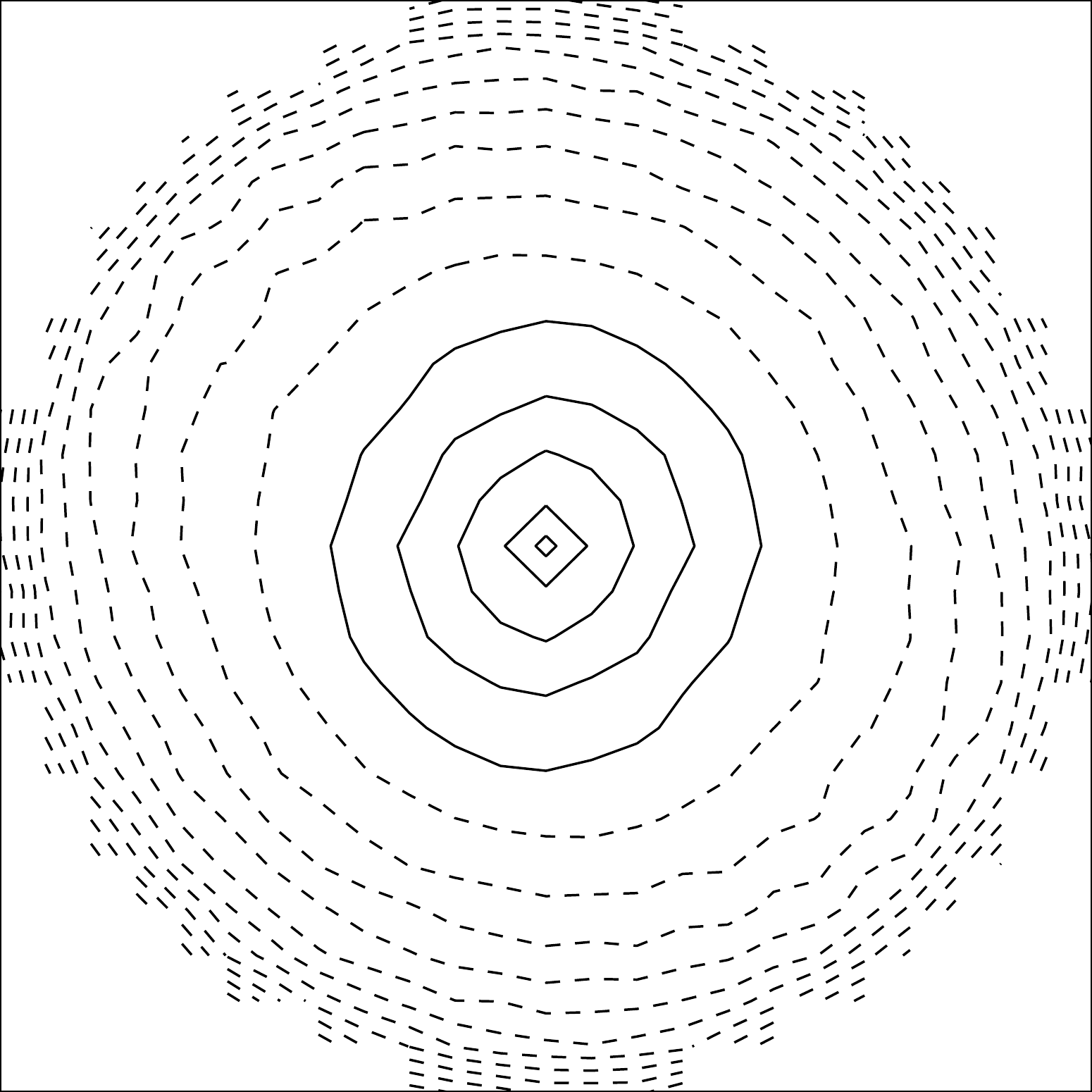}
  \caption[result 6941 (ASW000102p)]{A simulated lens that mimics a
    lensed quasar, and model results.  The left panels derive from the
    simulation, and the right panels are \spl output.  Details of
    individual panels are in \secref{example_models}.}

  \label{fig:6941}
\end{figure}

\begin{figure}
  \centering

  \includegraphics[width=\myplotswidth]{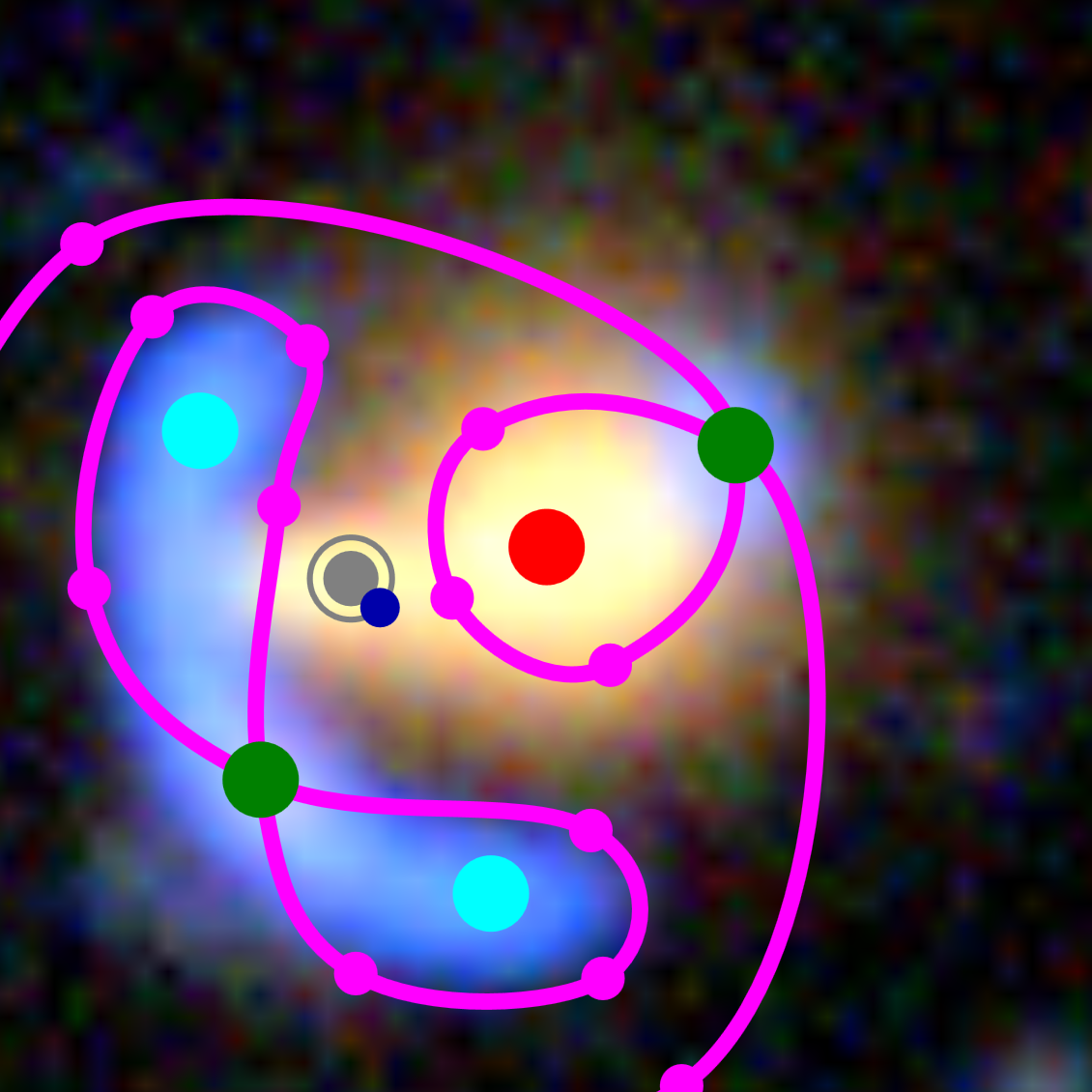}
  \includegraphics[width=\myplotswidth]{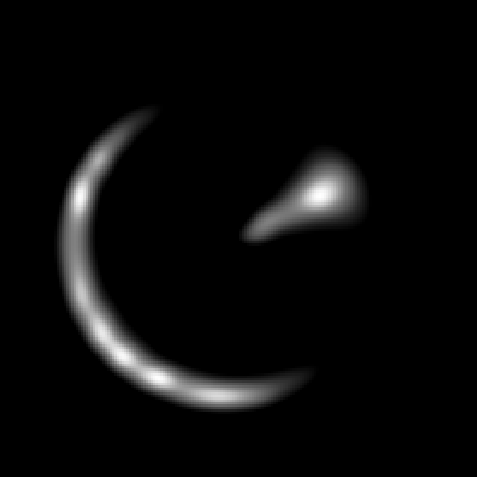} \\
  \includegraphics[width=\myplotswidth]{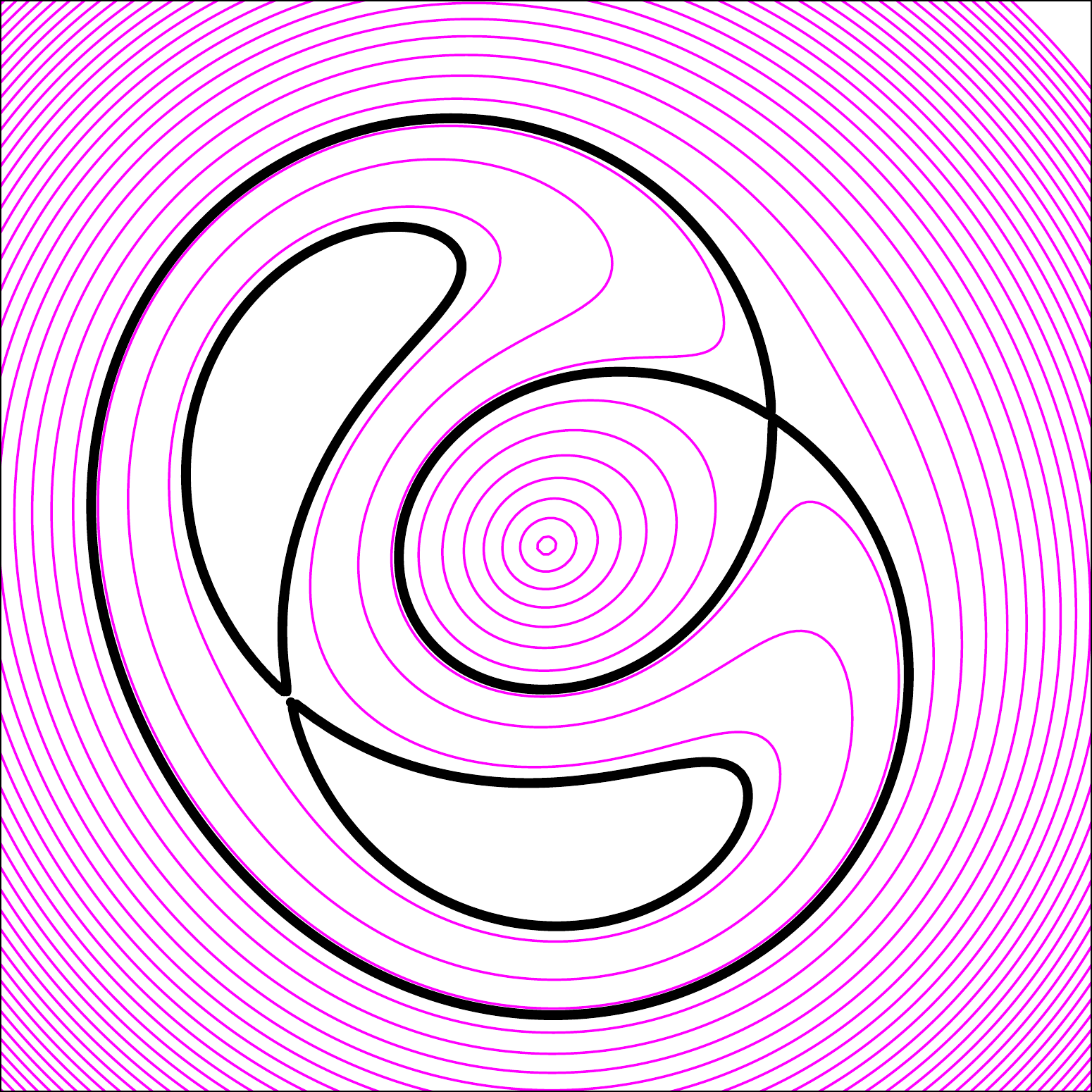}
  \includegraphics[width=\myplotswidth]{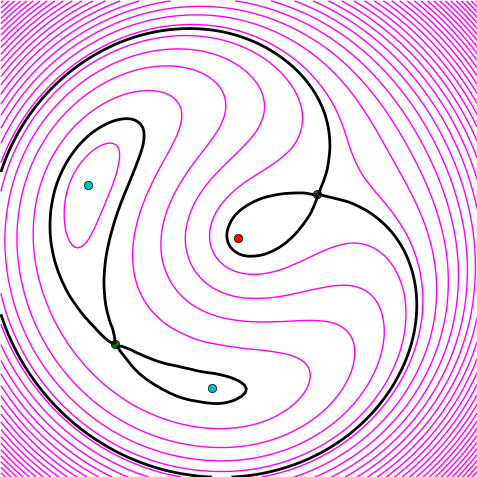} \\
  \includegraphics[width=\myplotswidth]{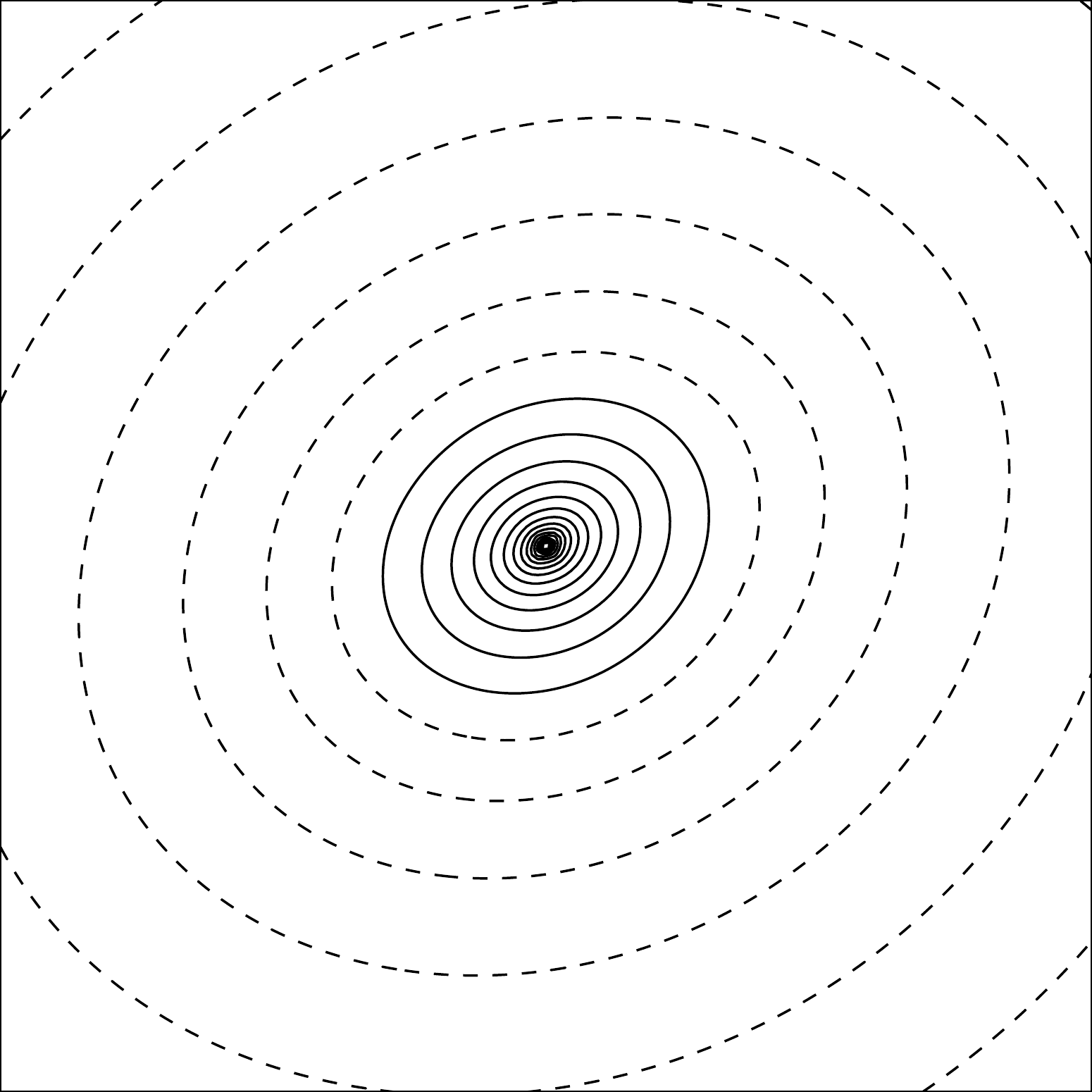}
  \includegraphics[width=\myplotswidth]{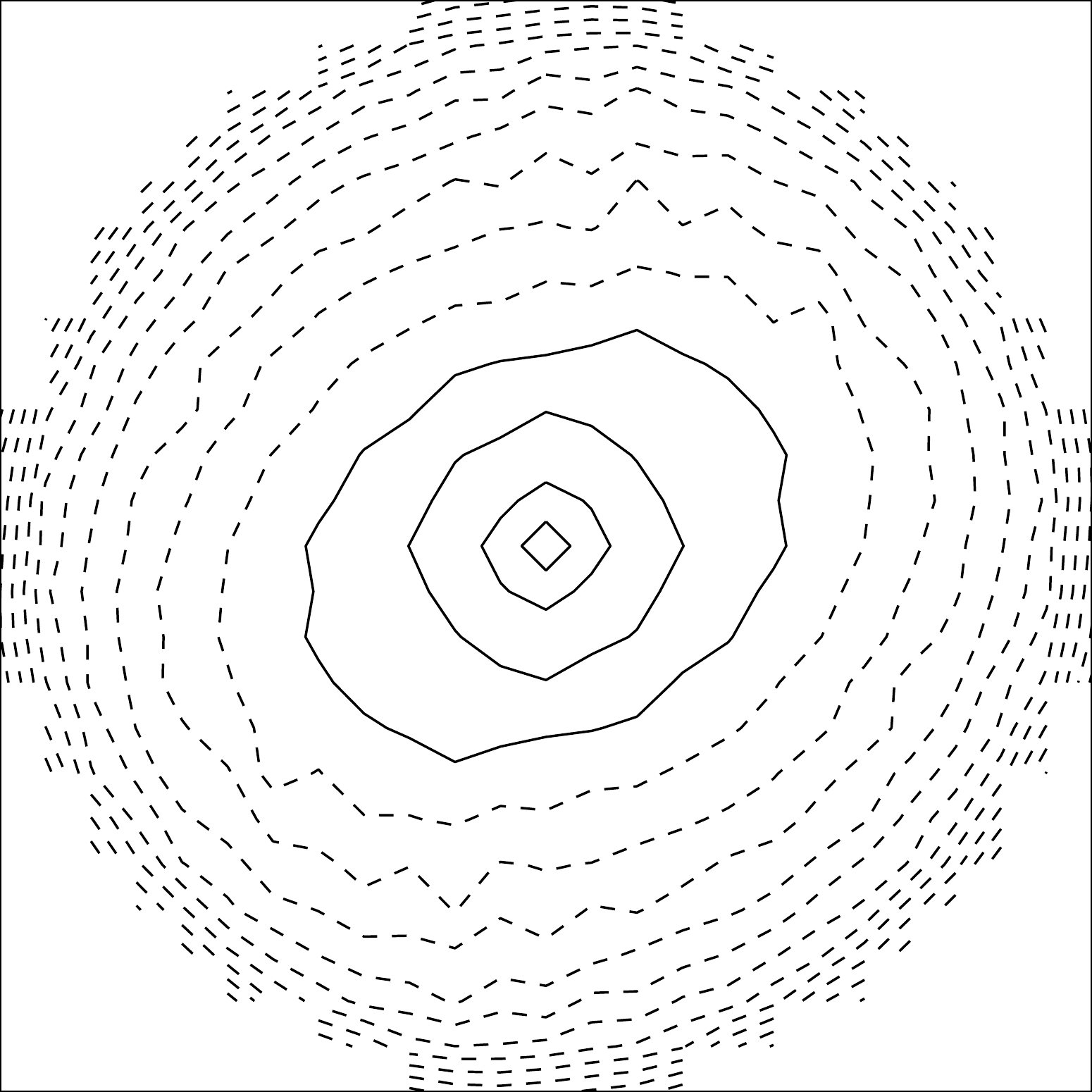}

  \caption[result 6990 (ASW0004oux)]{Results from a system with an arc
    plus a counter-image, typical of lensed galaxies. (See Section
    \ref{sec:example_models} for details.)}
  \label{fig:6990}
\end{figure}

\begin{figure}
  \centering

  \includegraphics[width=\myplotswidth]{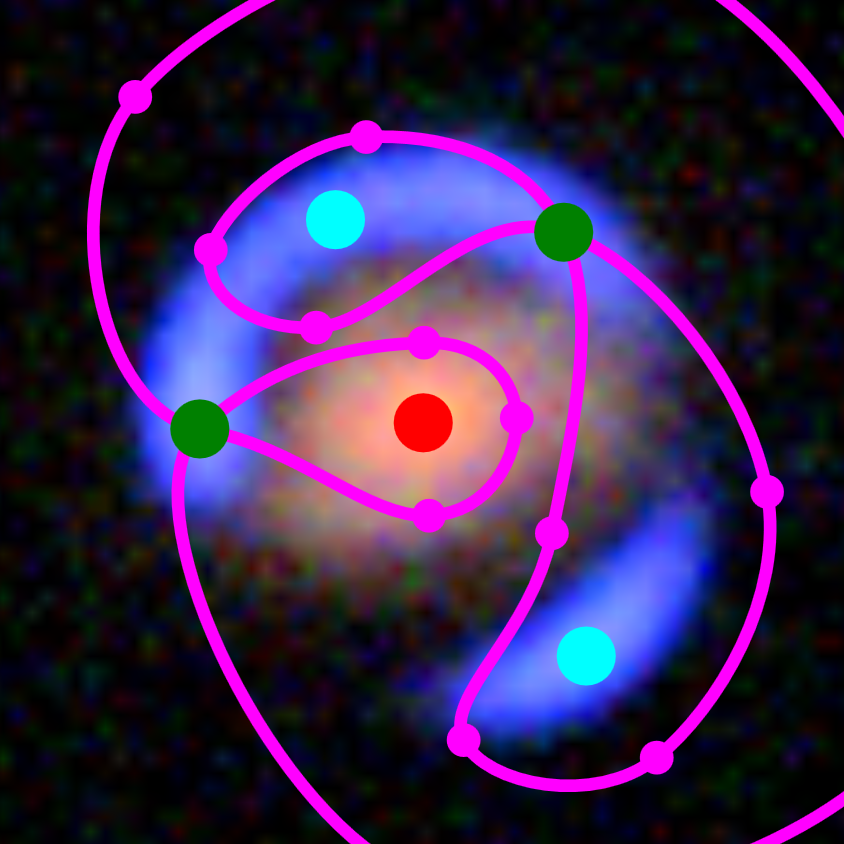}
  \includegraphics[width=\myplotswidth]{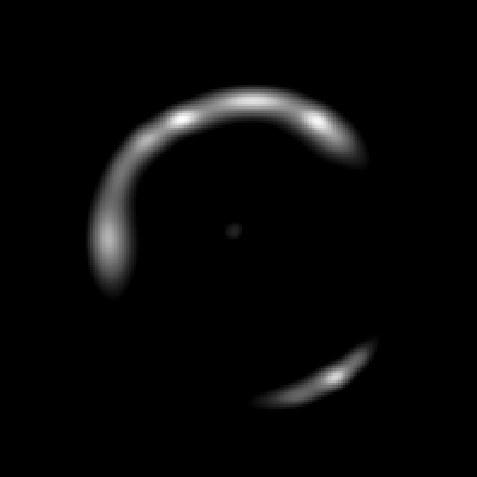} \\
  \includegraphics[width=\myplotswidth]{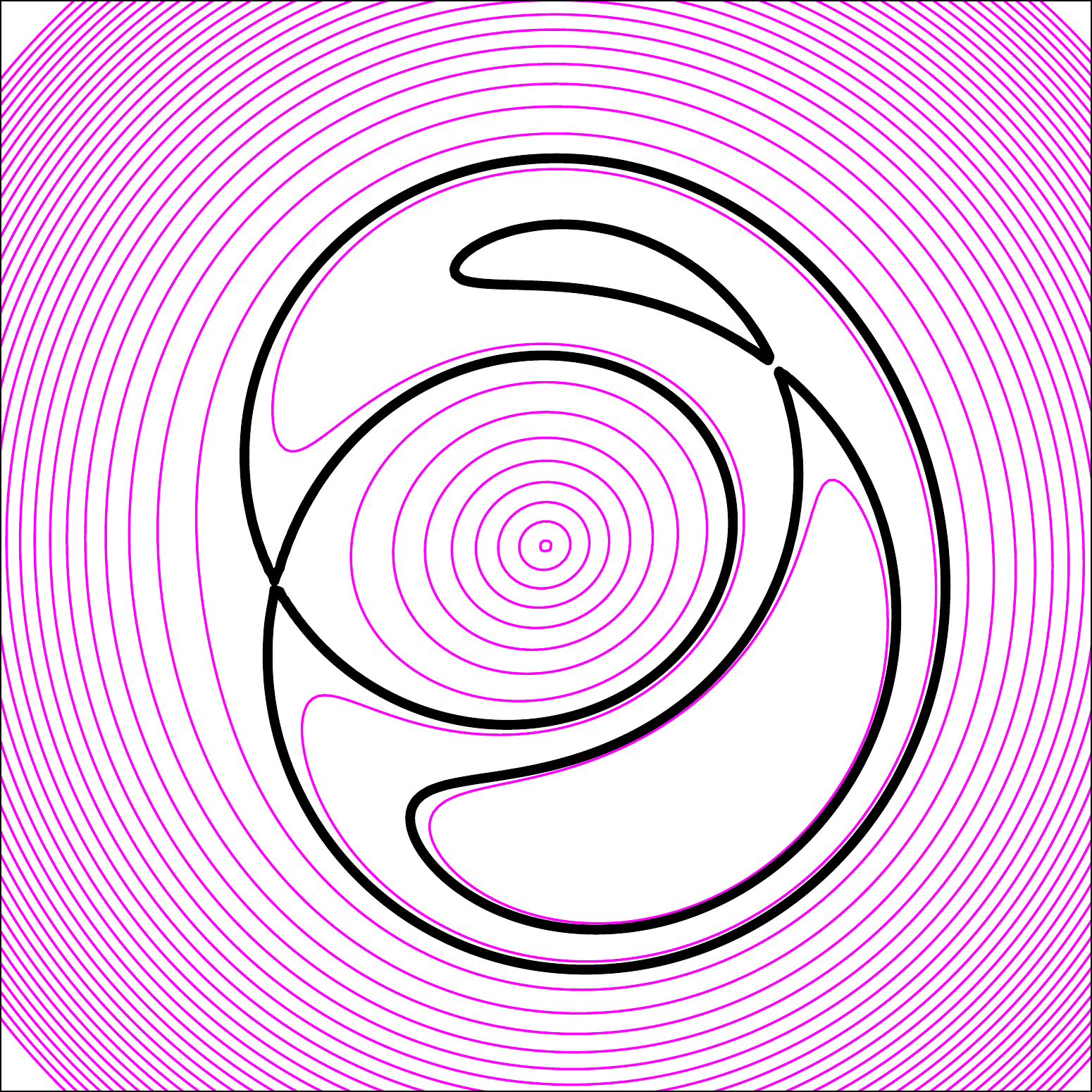}
  \includegraphics[width=\myplotswidth]{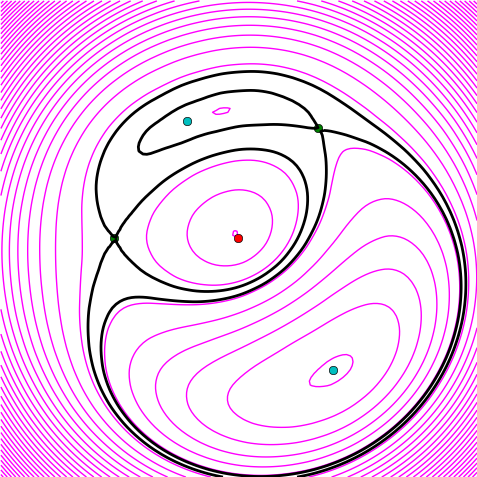} \\
  \includegraphics[width=\myplotswidth]{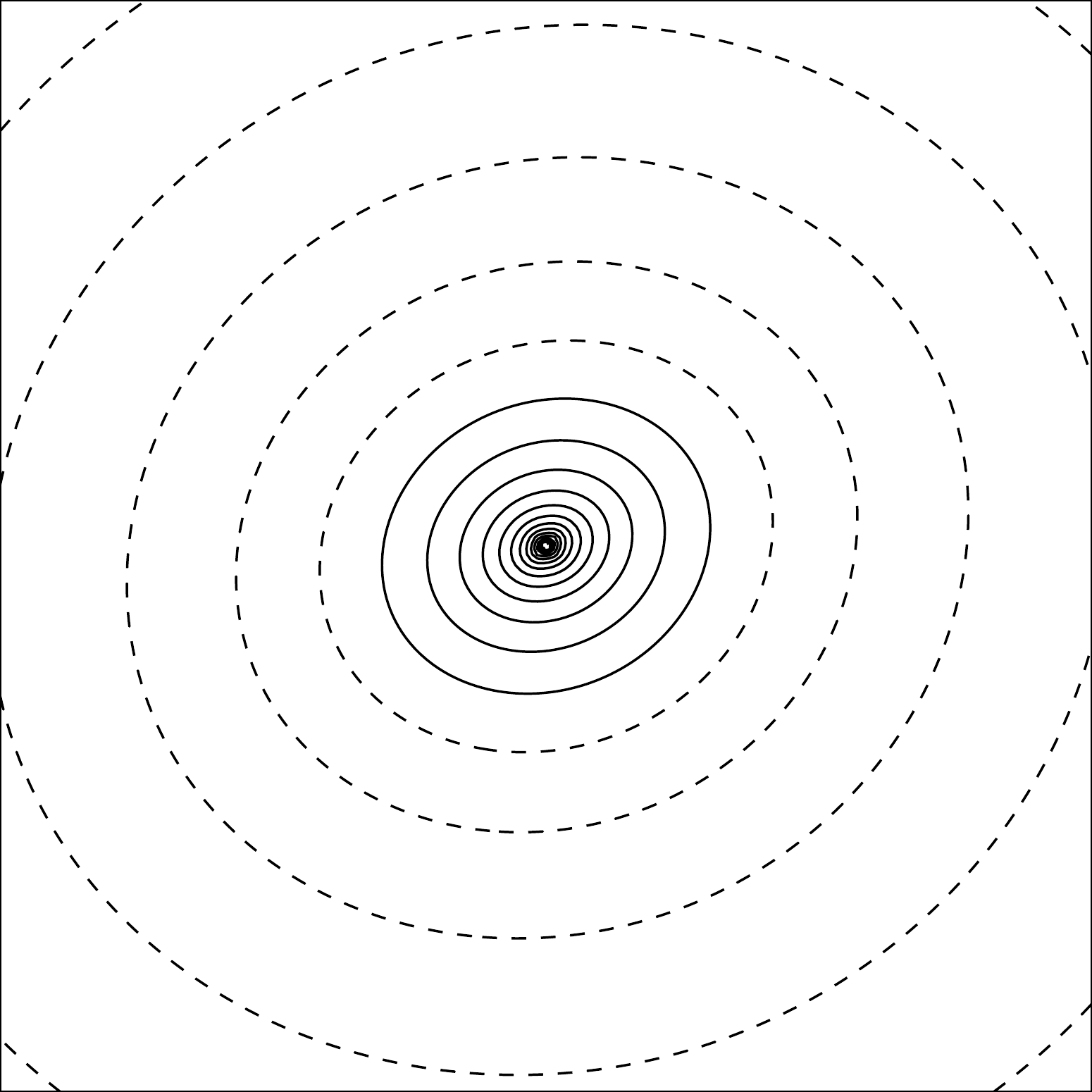}
  \includegraphics[width=\myplotswidth]{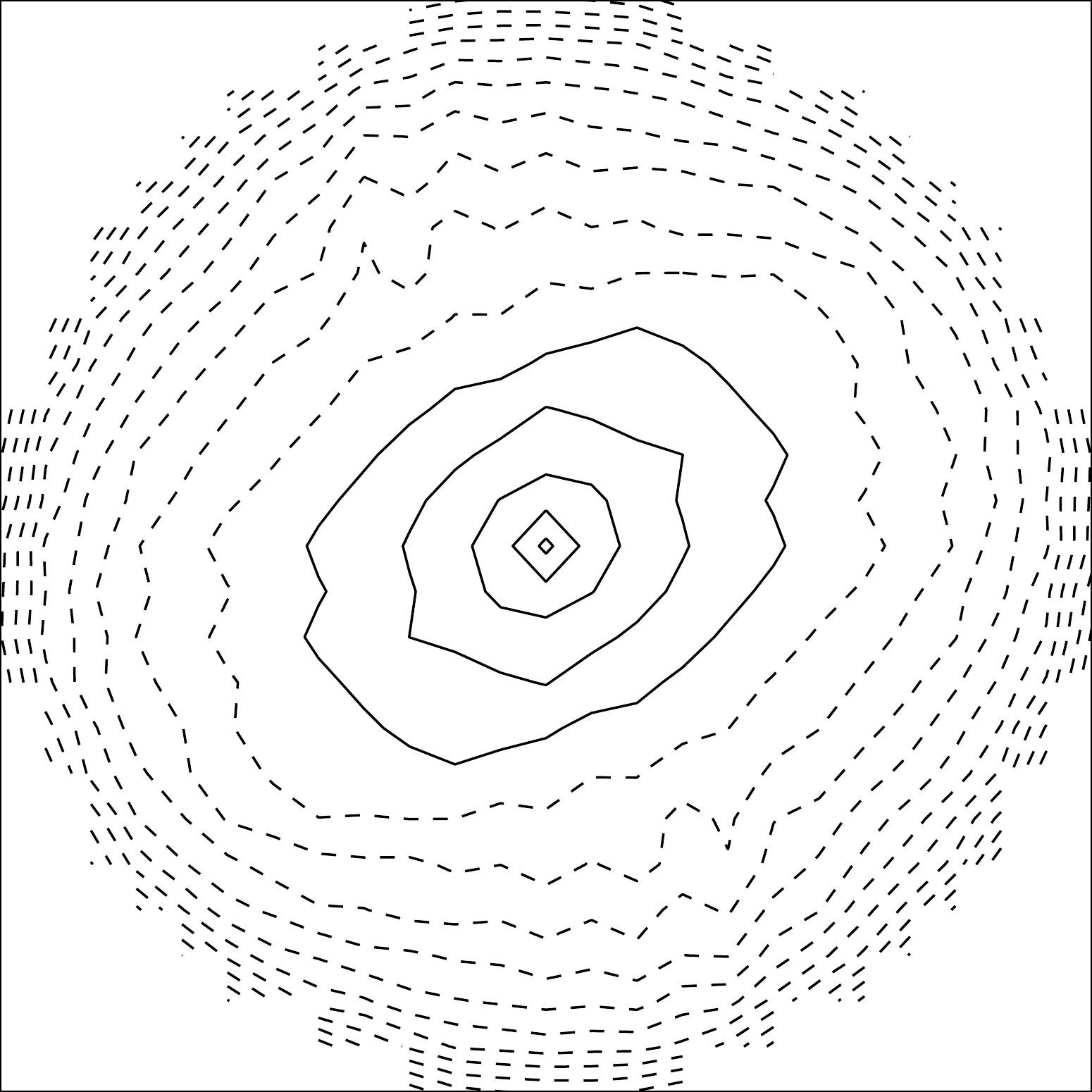}

  \caption[result 6919 (ASW0002z6f)]{Another configuration of arc plus
    counter-image: an arc and counter image where the arc is closer to
    the lensing galaxy than the counter image. (See
    \secref{example_models} for details.)
    \vspace{1em}}
  \label{fig:6919}
\end{figure}

\begin{figure}
  \centering
  \includegraphics[width=\myplotswidth]{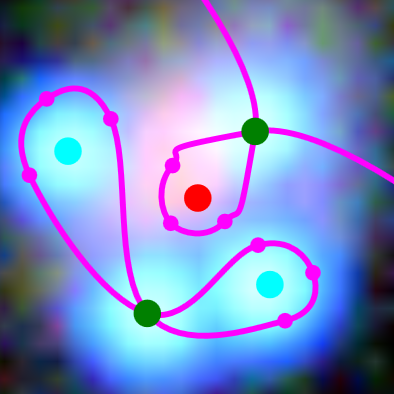}
  \includegraphics[width=\myplotswidth]{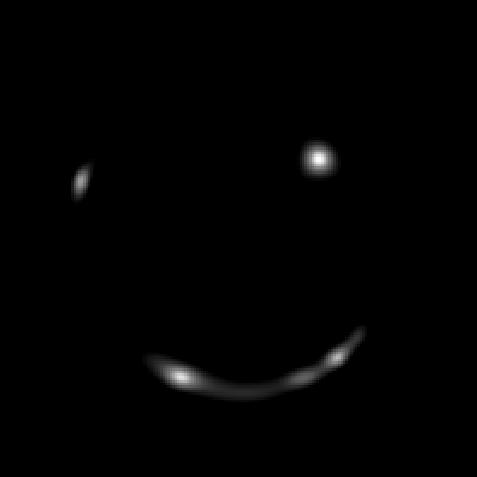} \\
  \includegraphics[width=\myplotswidth]{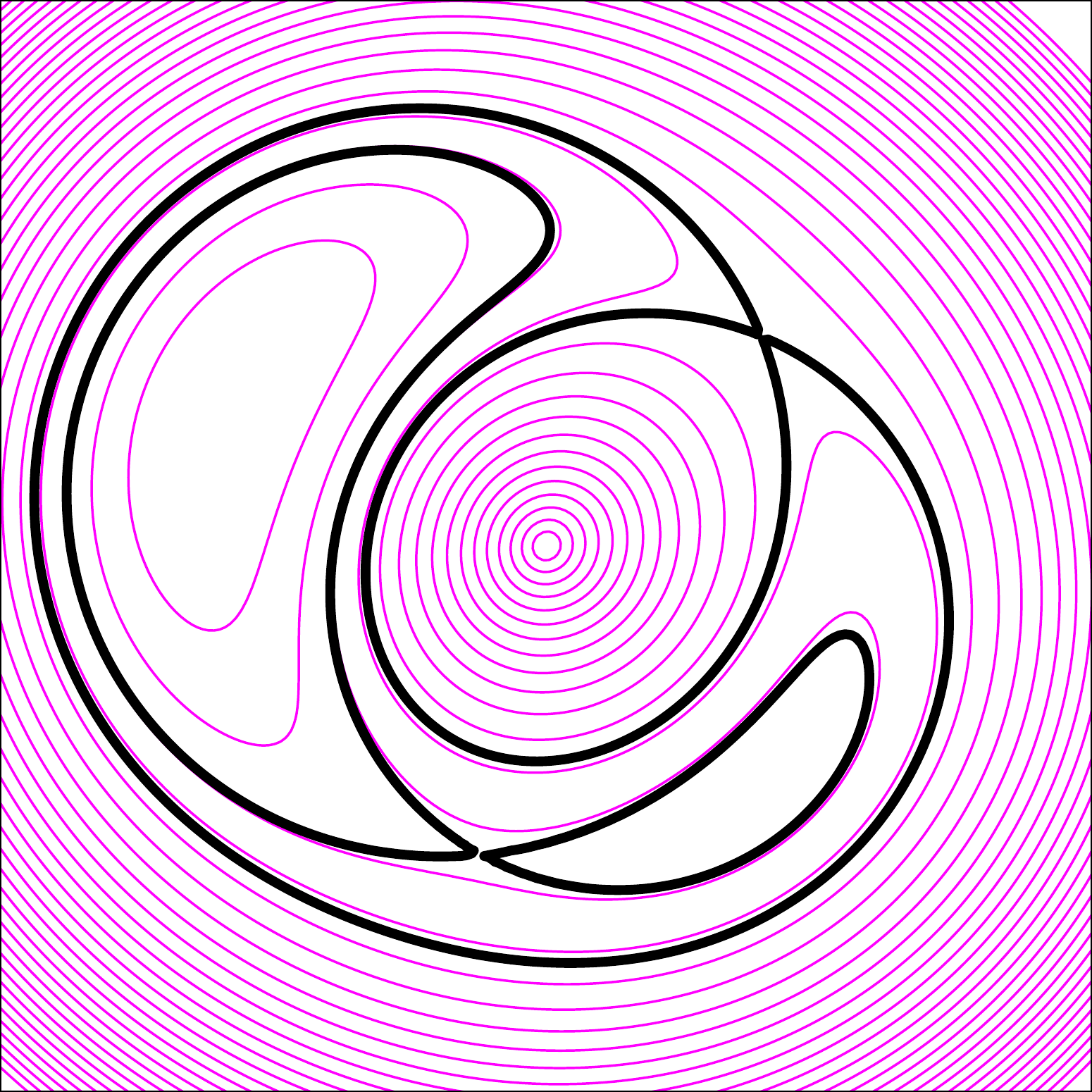}
  \includegraphics[width=\myplotswidth]{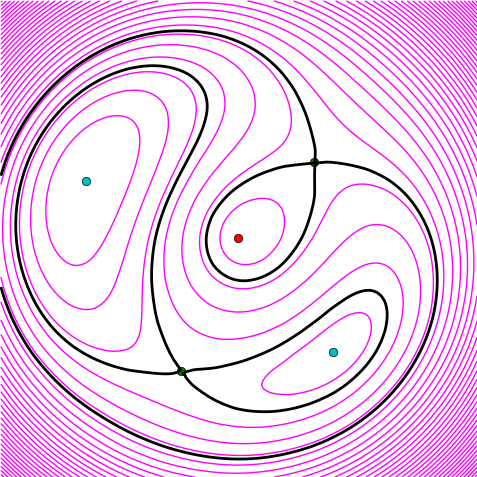} \\
  \includegraphics[width=\myplotswidth]{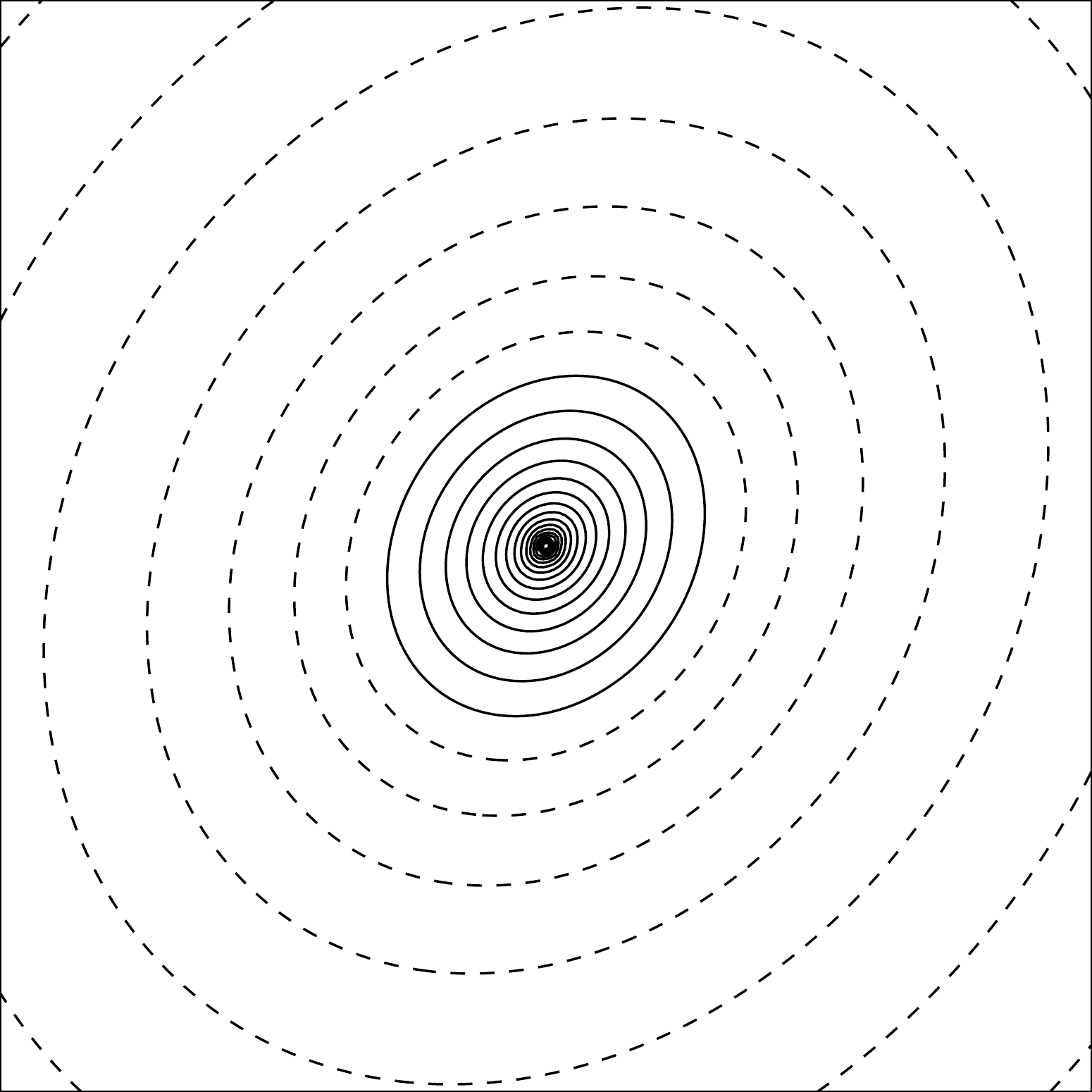}
  \includegraphics[width=\myplotswidth]{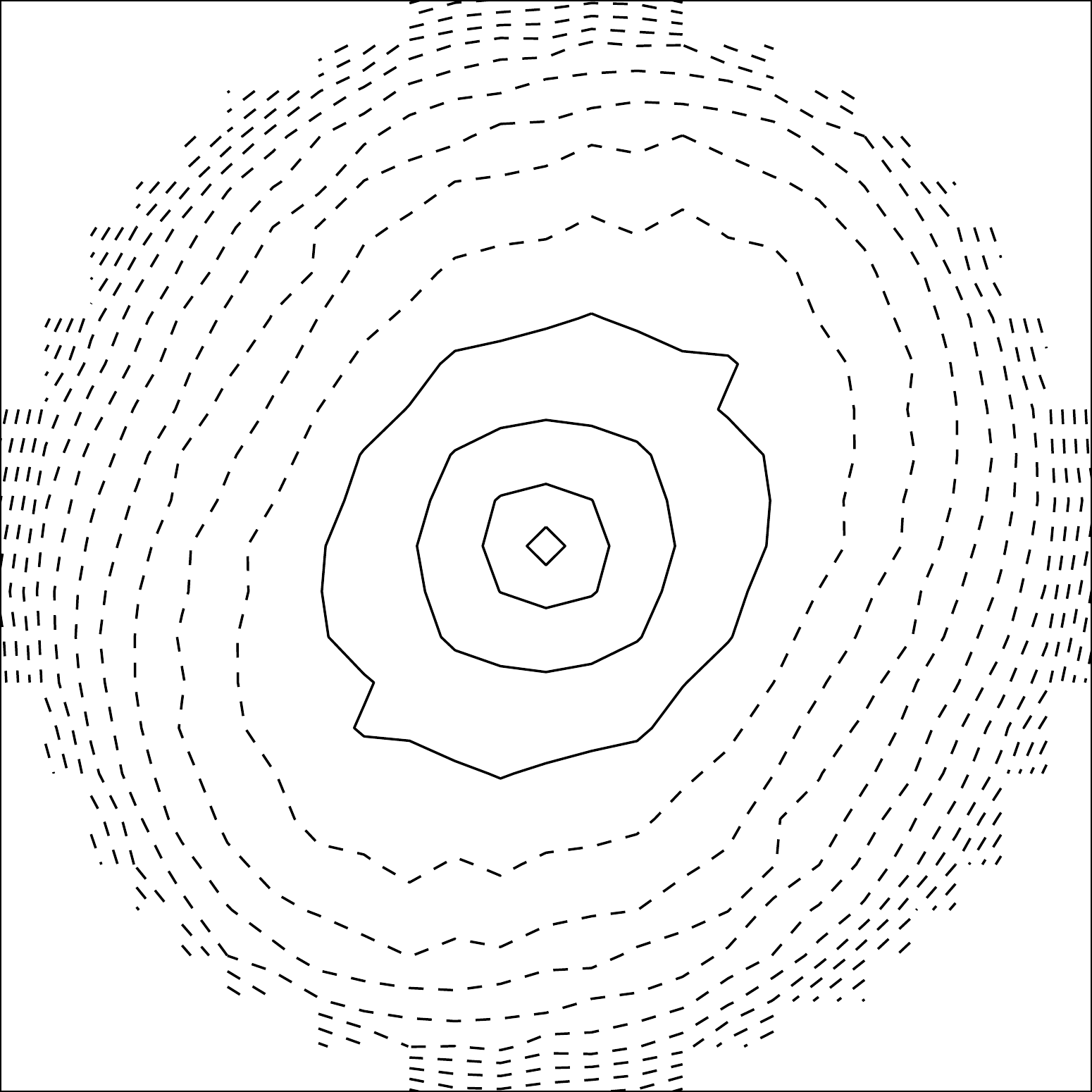}

  \caption[result 6915 (ASW0001hpf)]{A four-image configuration
    typical of lensed quasars. (See \secref{example_models} for
    details.)}
  \label{fig:6915}
\end{figure}

\begin{figure}
  \centering
  \includegraphics[width=\myplotswidth]{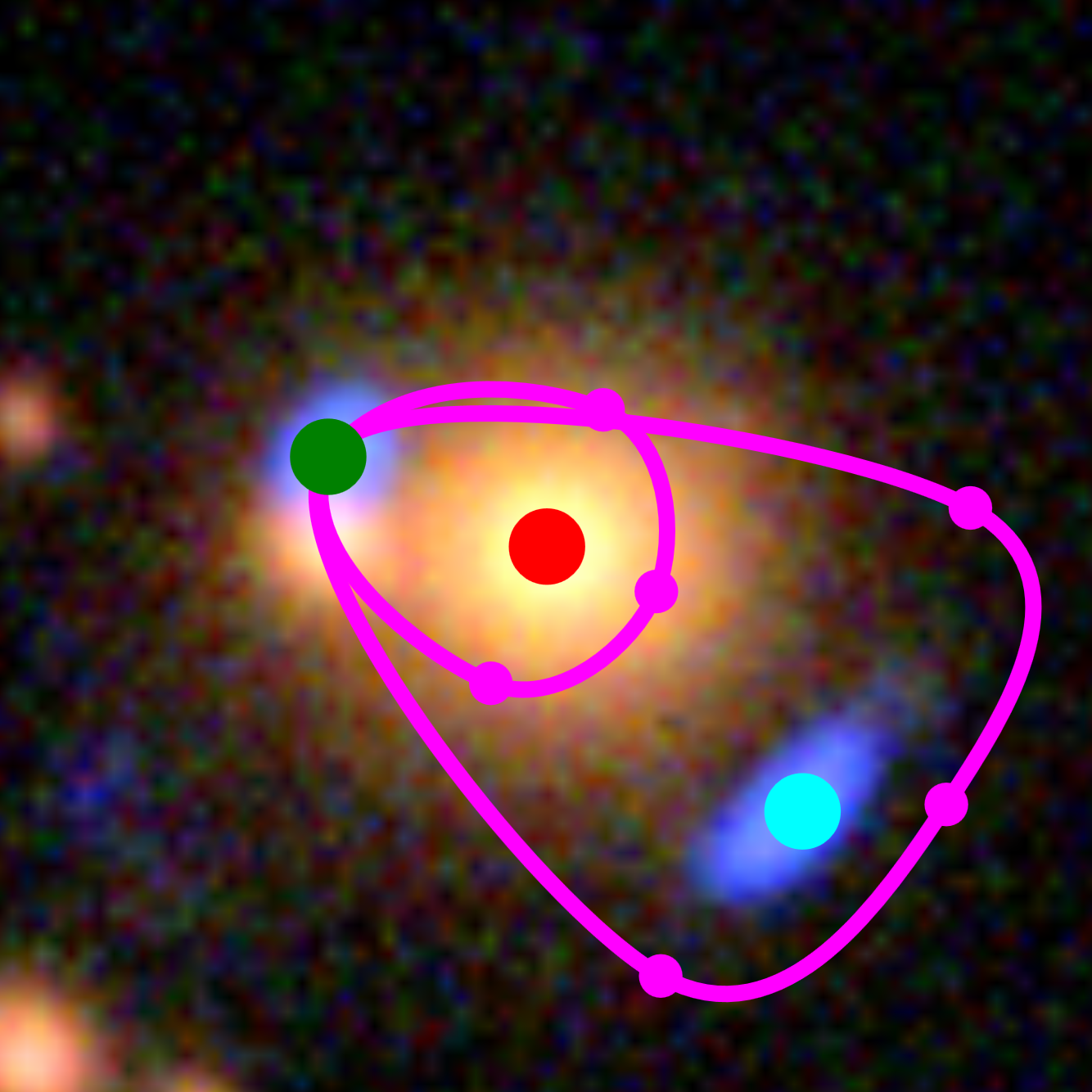}
  \includegraphics[width=\myplotswidth]{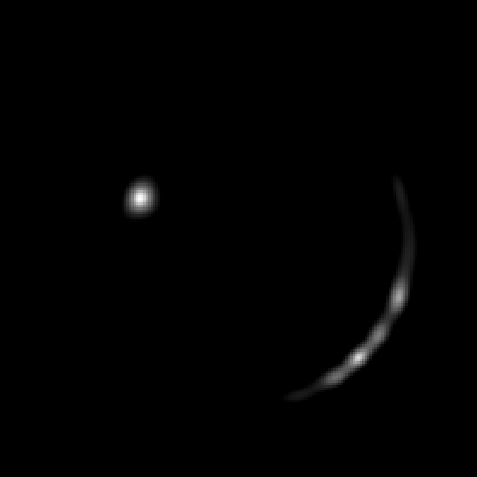} \\
  \includegraphics[width=\myplotswidth]{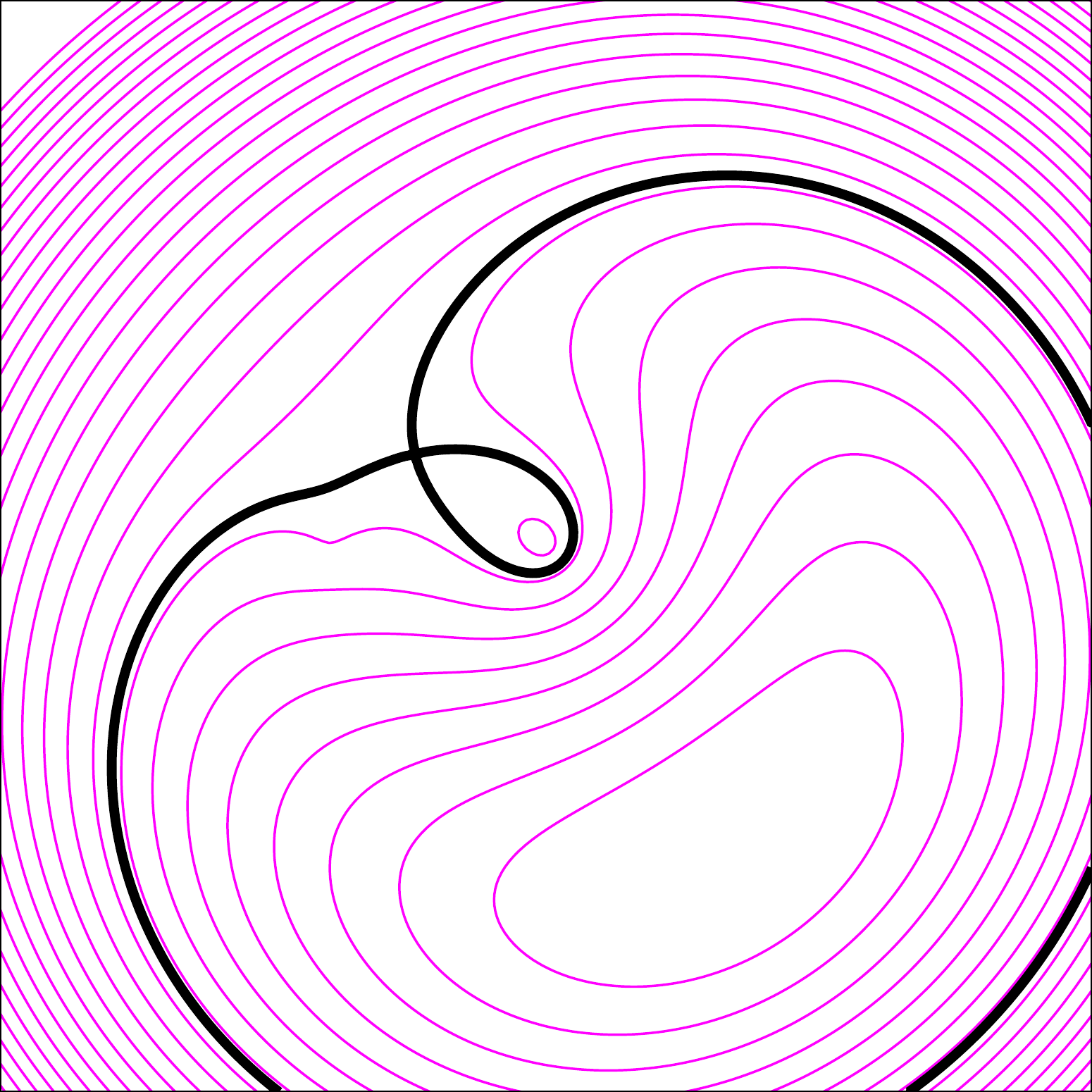}
  \includegraphics[width=\myplotswidth]{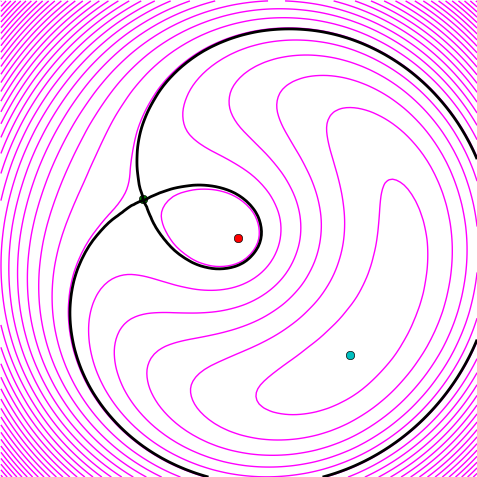} \\
  \includegraphics[width=\myplotswidth]{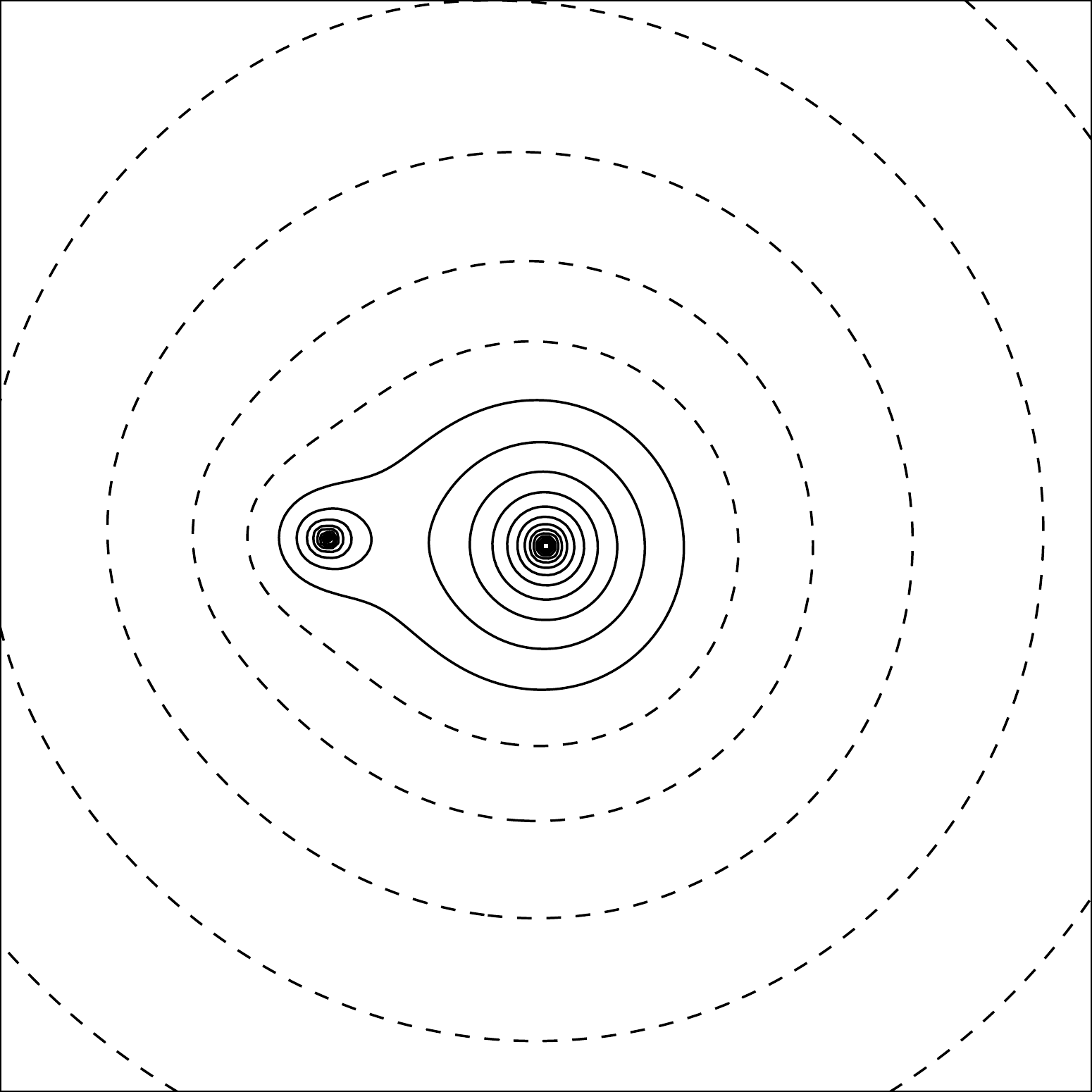}
  \includegraphics[width=\myplotswidth]{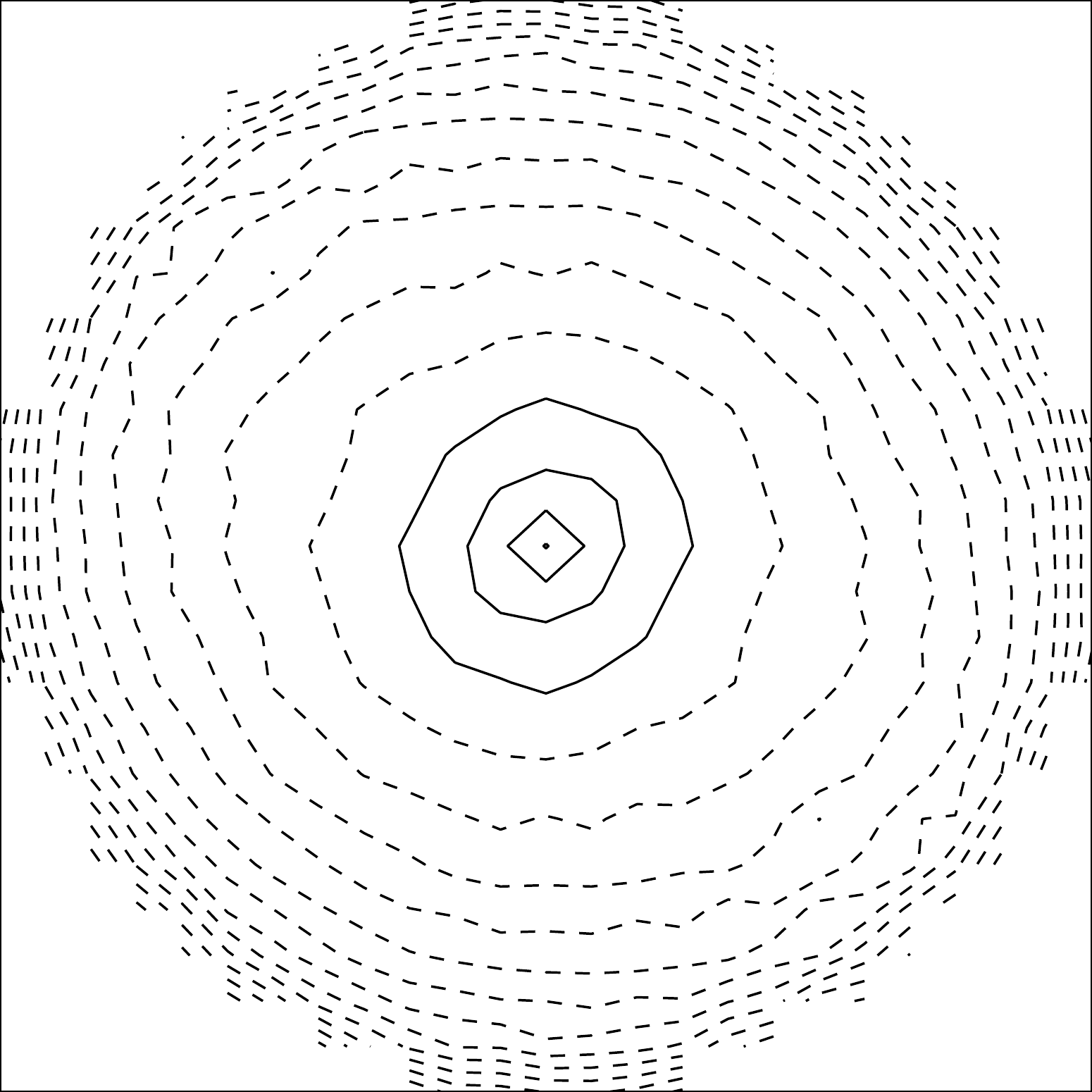}
  \caption[result 6975 (ASW000195x)]{A lens with unrecovered mass
    substructure. (See \secref{example_models} for details.)
    \vspace{1em}
    }
  \label{fig:6975}
\end{figure}

\begin{figure}
  \centering

  \includegraphics[width=\myplotswidth]{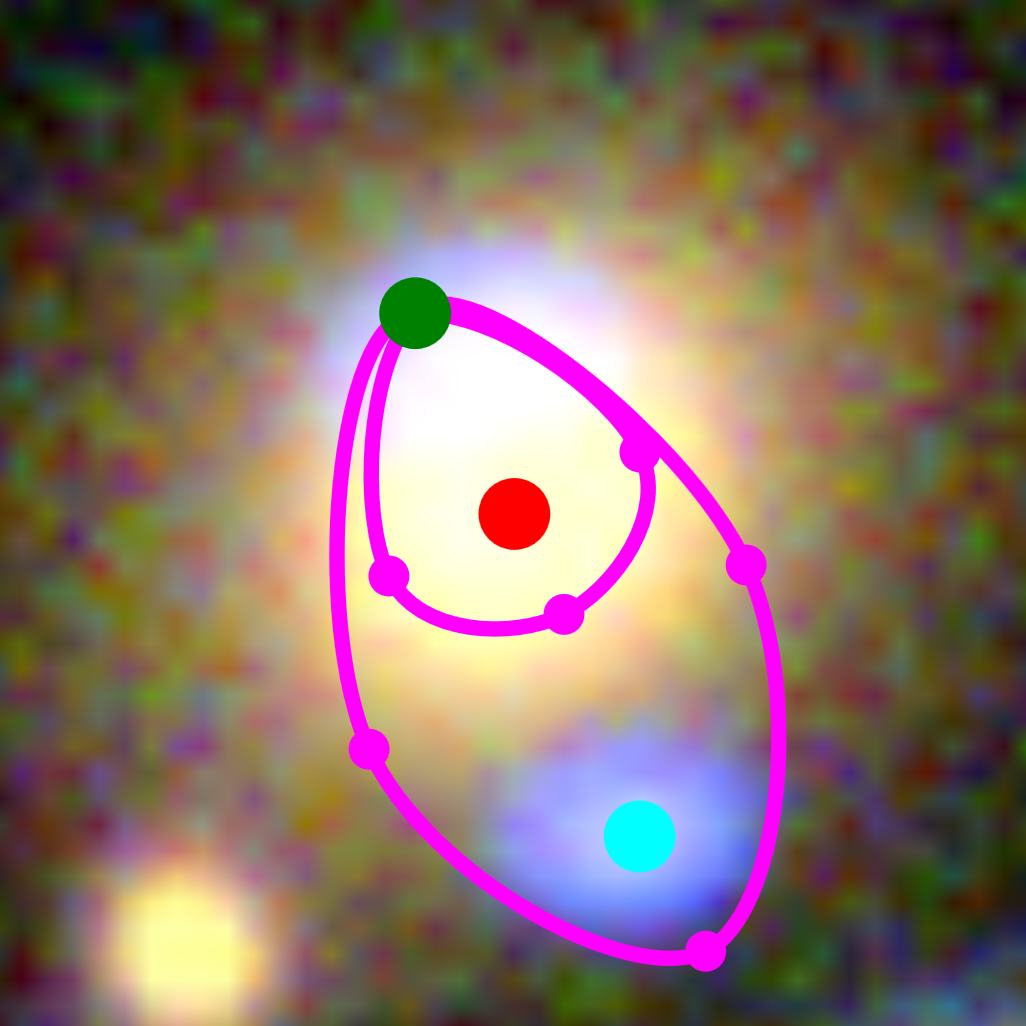}
  \includegraphics[width=\myplotswidth]{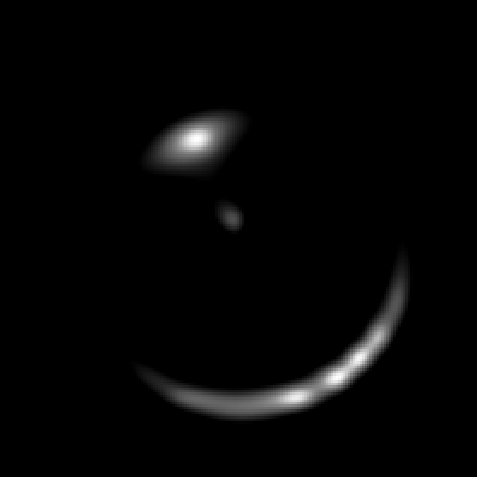} \\
  \includegraphics[width=\myplotswidth]{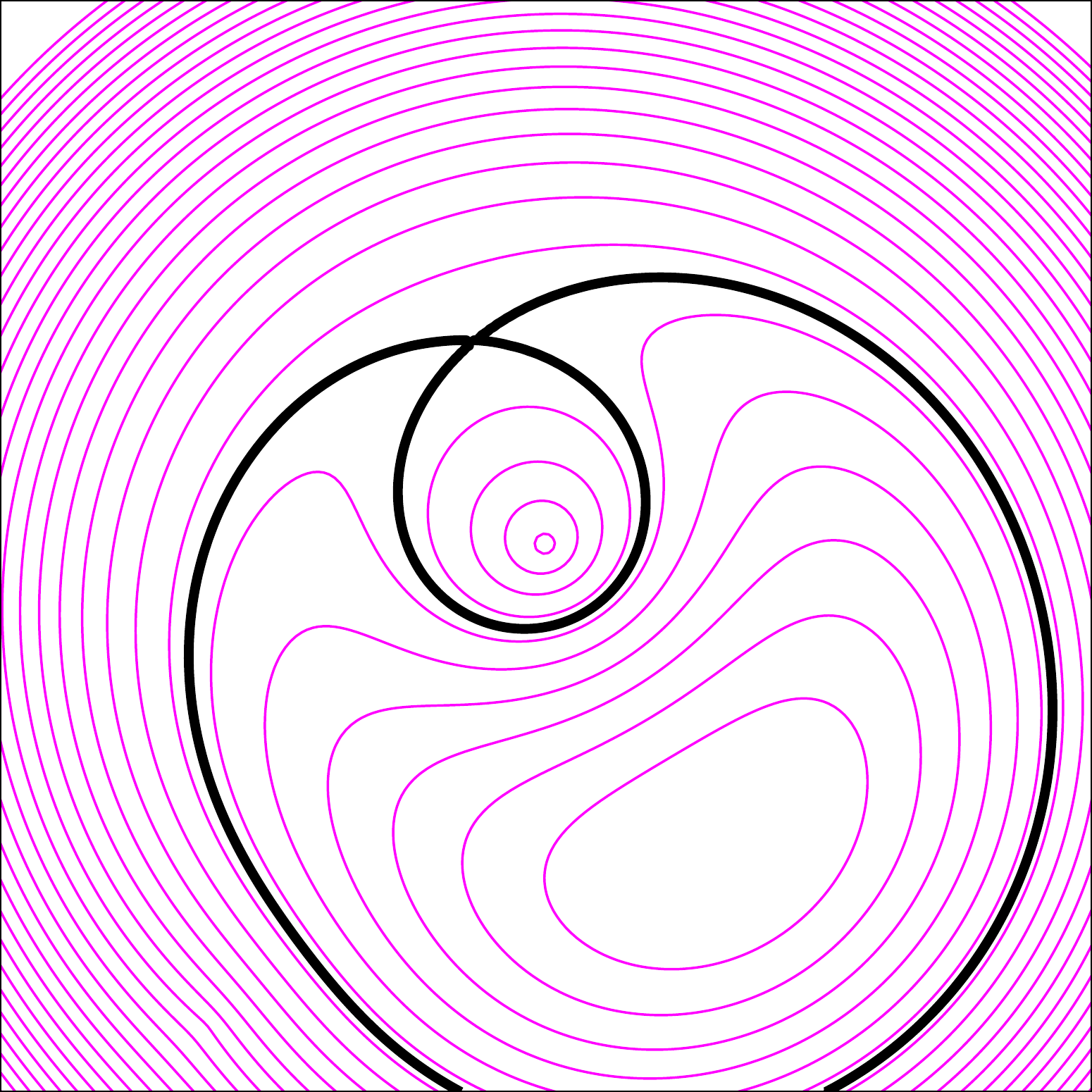}
  \includegraphics[width=\myplotswidth]{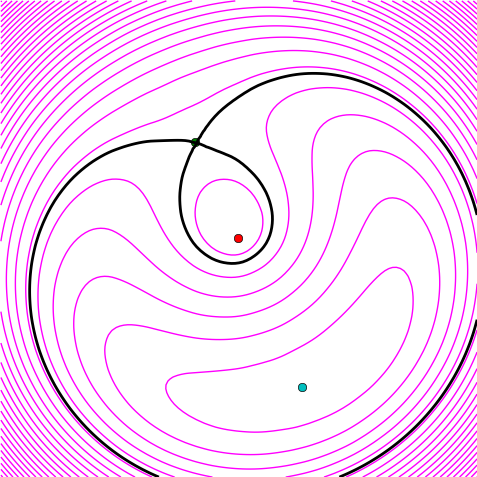} \\
  \includegraphics[width=\myplotswidth]{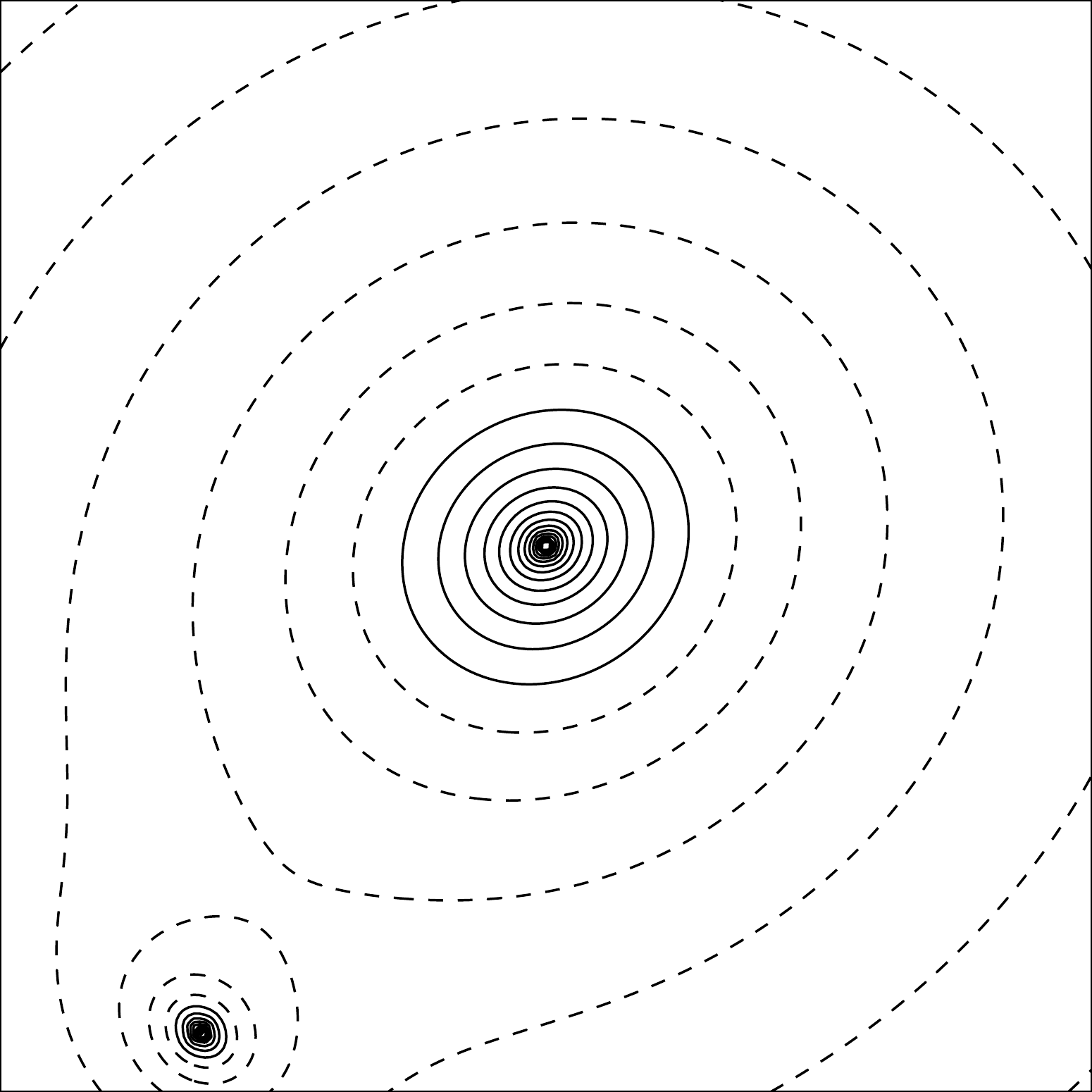}
  \includegraphics[width=\myplotswidth]{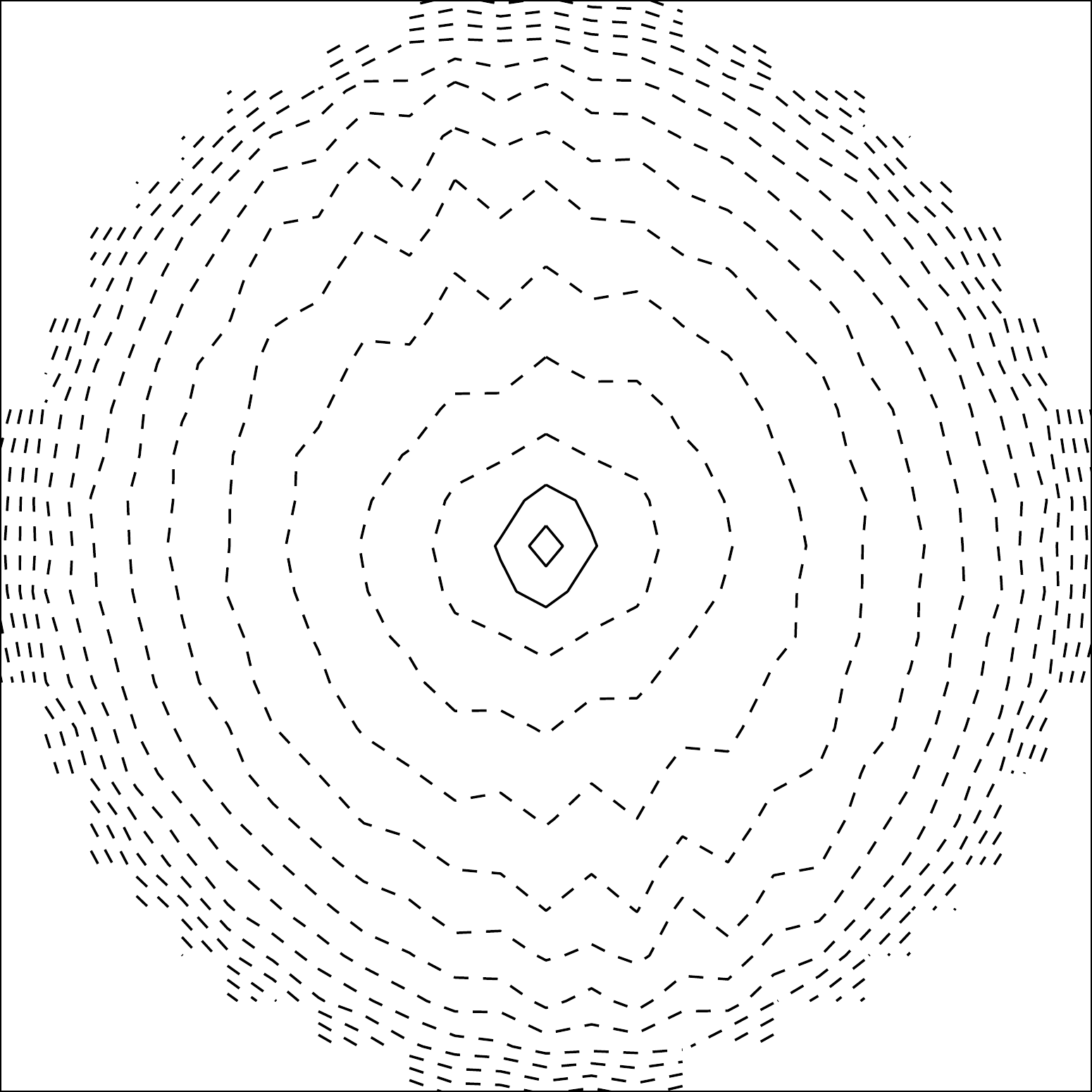}

  \caption[result 6937 (ASW0000vqg)]{A sim with unrecovered
    substructure, resulting in a poor mass model. (See Section
    \ref{sec:example_models} for details.)}
  \label{fig:6937}
\end{figure}

\begin{figure}
  \centering

  \includegraphics[width=\myplotswidth]{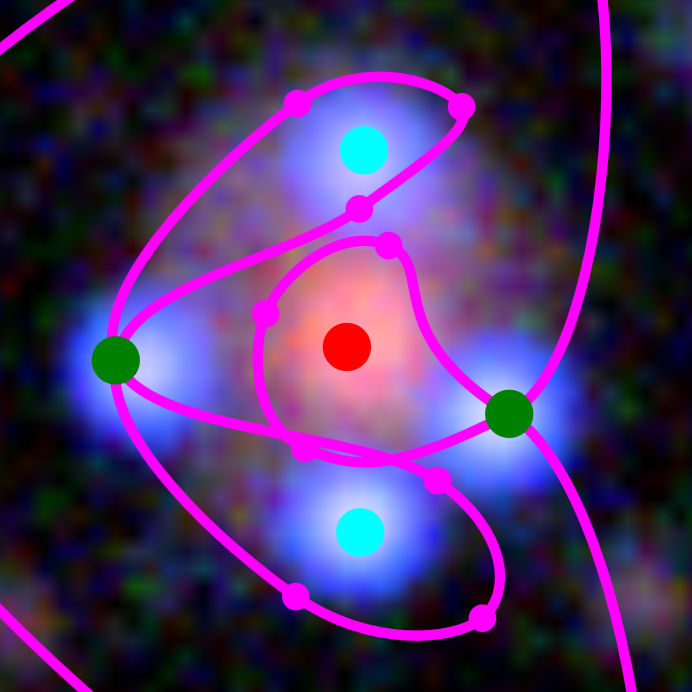}
  \includegraphics[width=\myplotswidth]{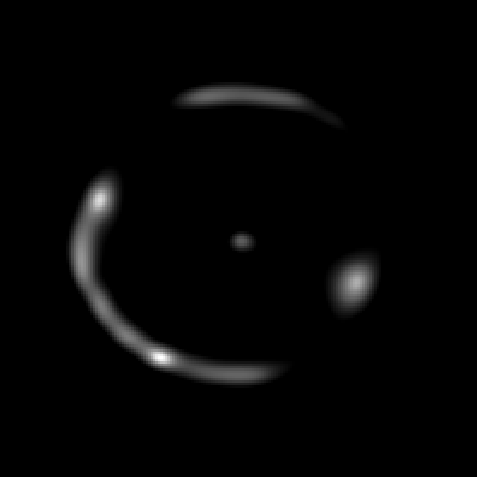} \\
  \includegraphics[width=\myplotswidth]{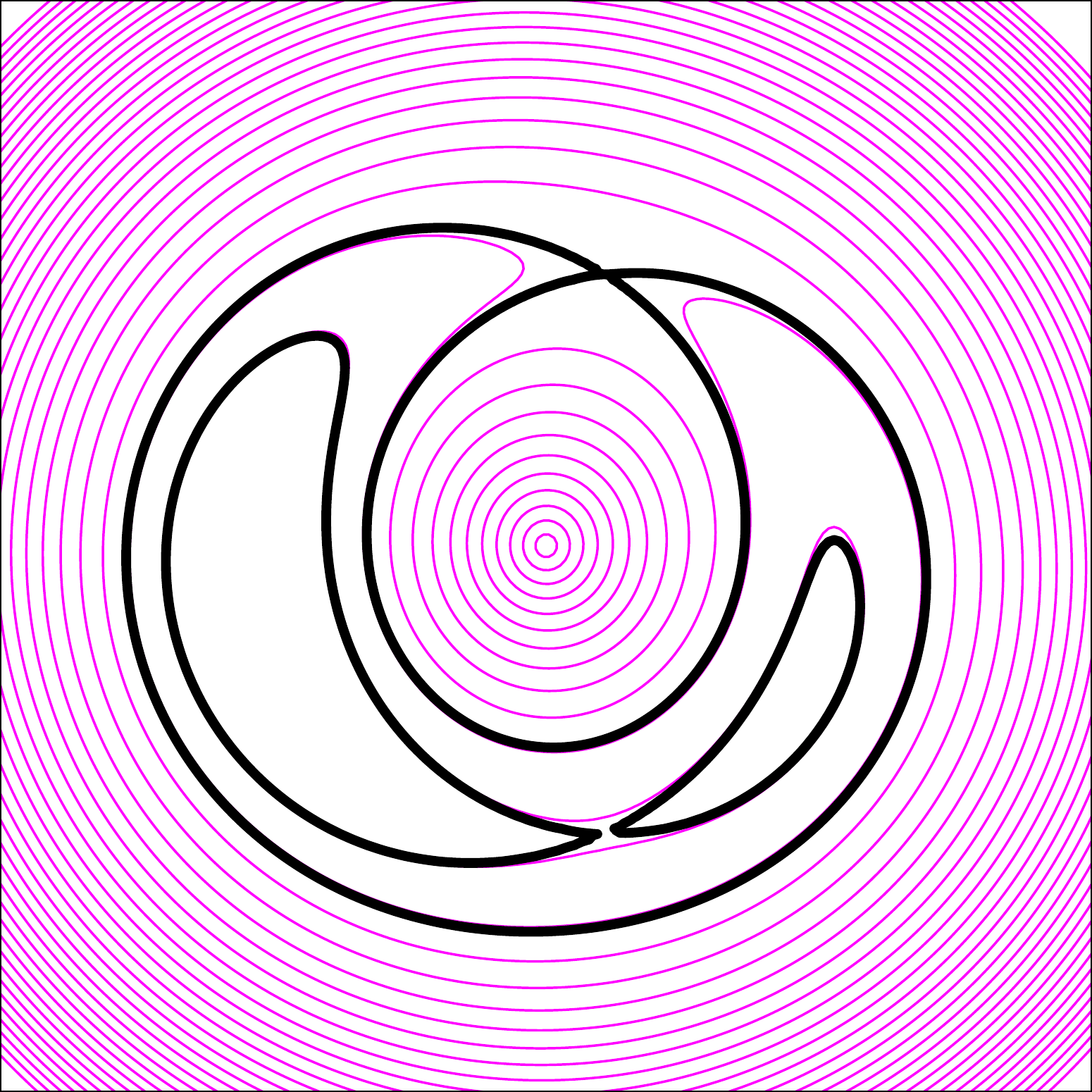}
  \includegraphics[width=\myplotswidth]{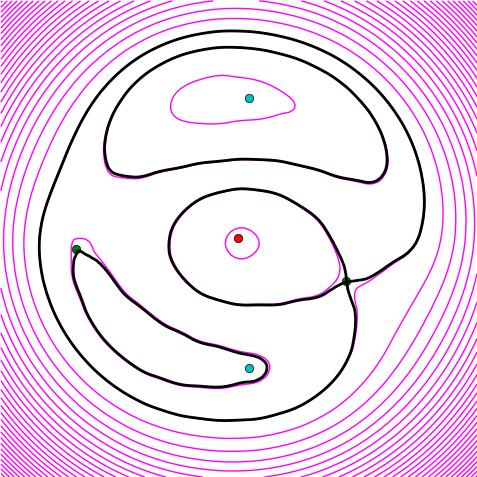} \\
  \includegraphics[width=\myplotswidth]{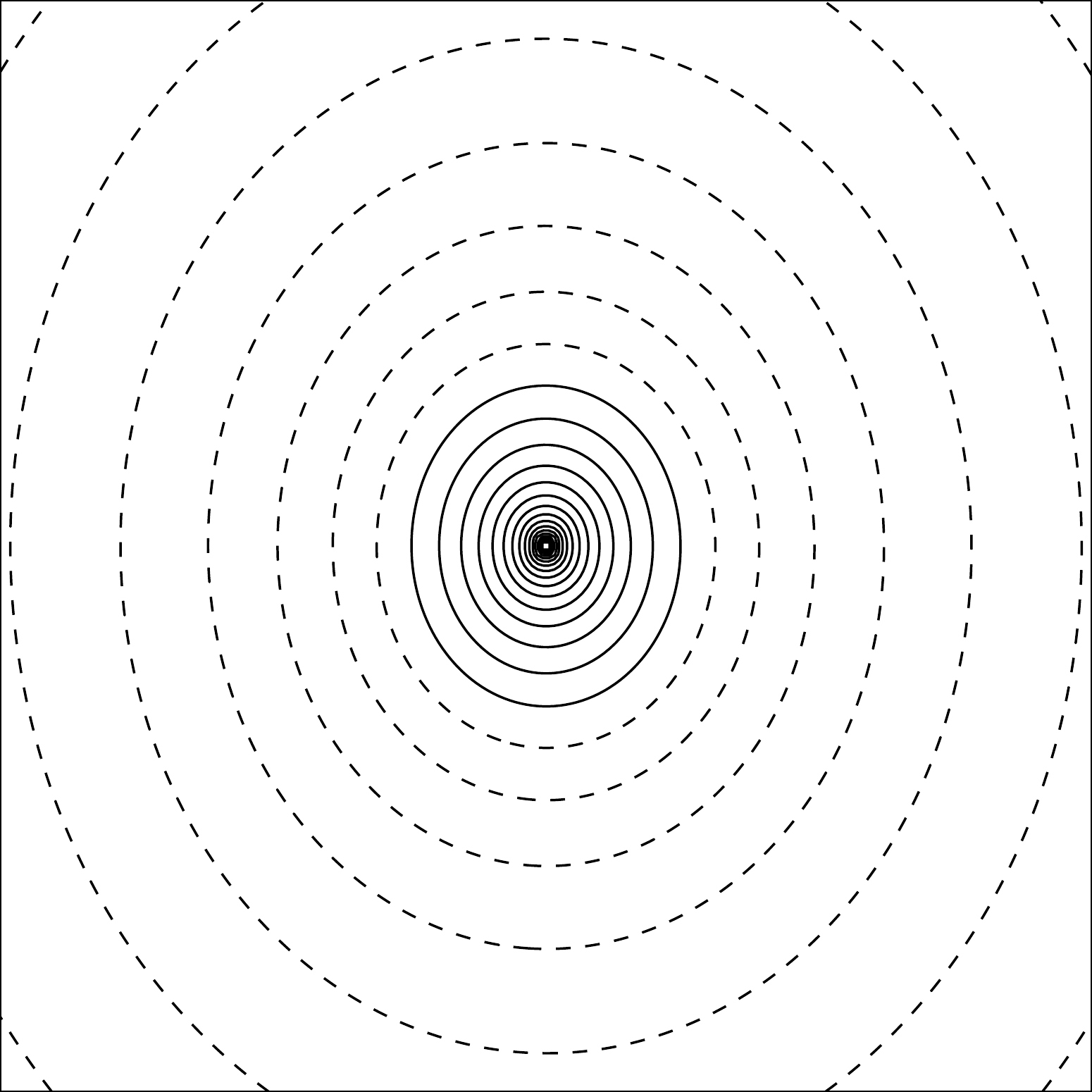}
  \includegraphics[width=\myplotswidth]{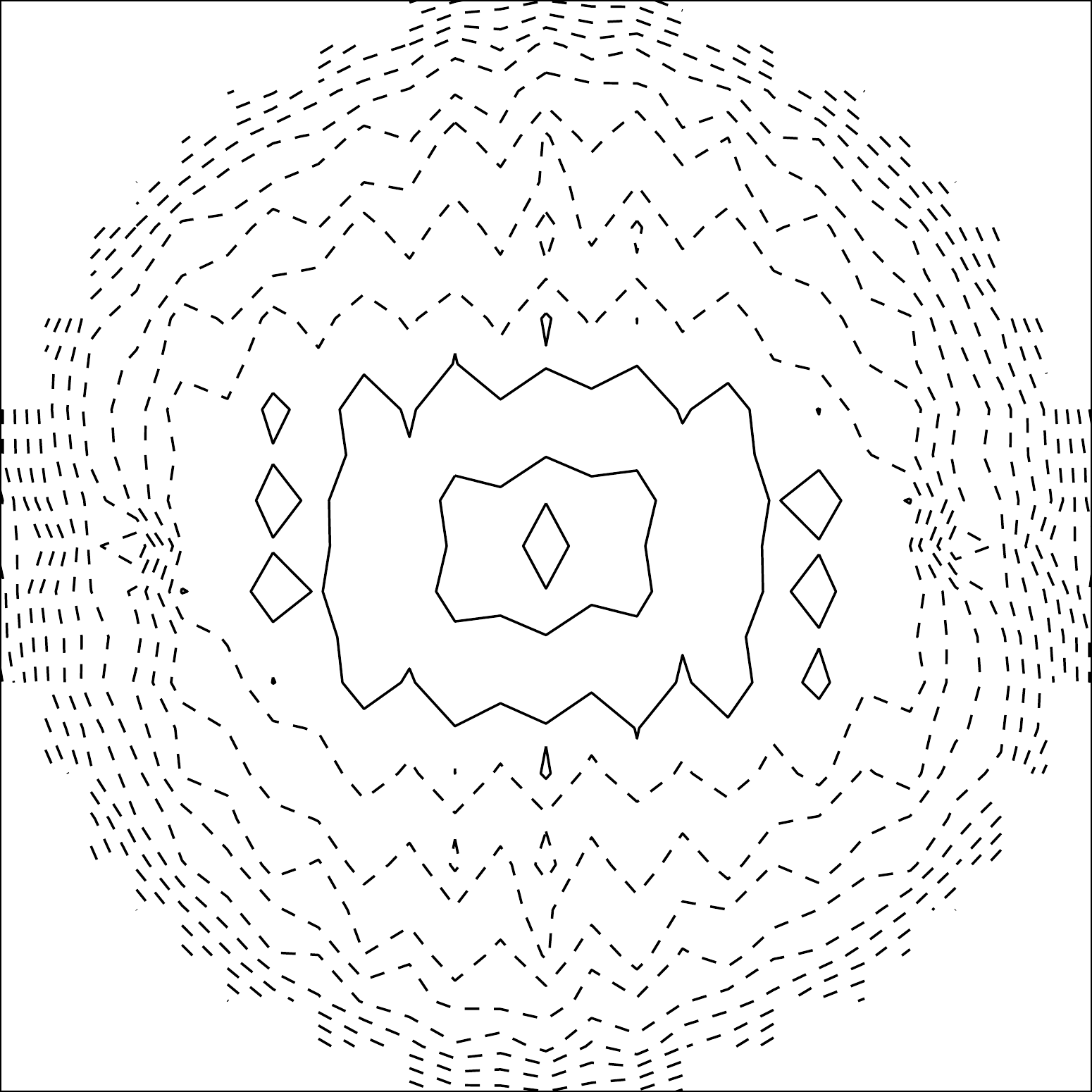}
  \caption[result 7025 (ASW0000h2m)]{A four-image system with image
    parities incorrectly identified.  The model is poor, but the
    estimated Einstein radius is not bad. (See Section
    \ref{sec:example_models} for details.)}
  \label{fig:7025}
\end{figure}

\begin{figure}
  \centering

  \includegraphics[width=\myplotswidth]{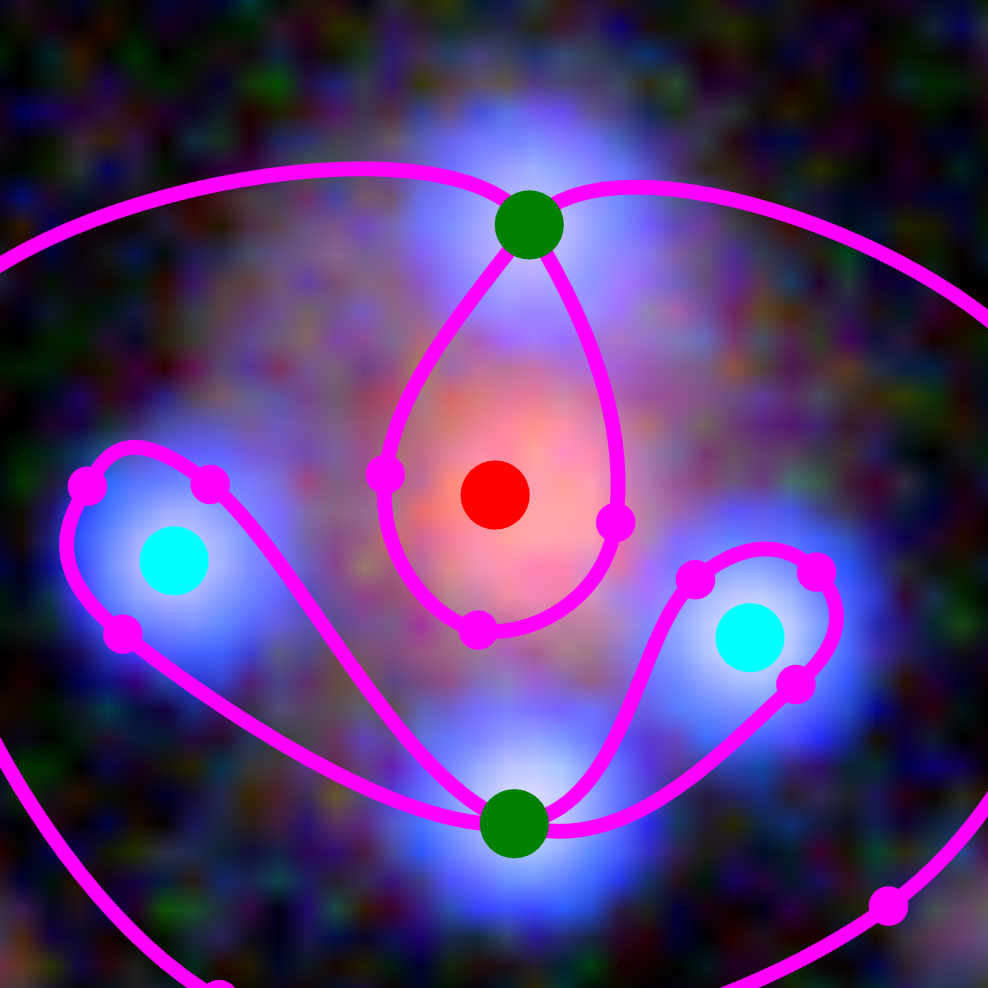}
  \includegraphics[width=\myplotswidth]{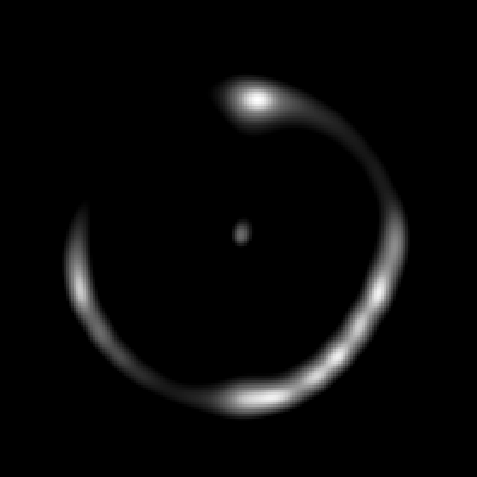} \\
  \includegraphics[width=\myplotswidth]{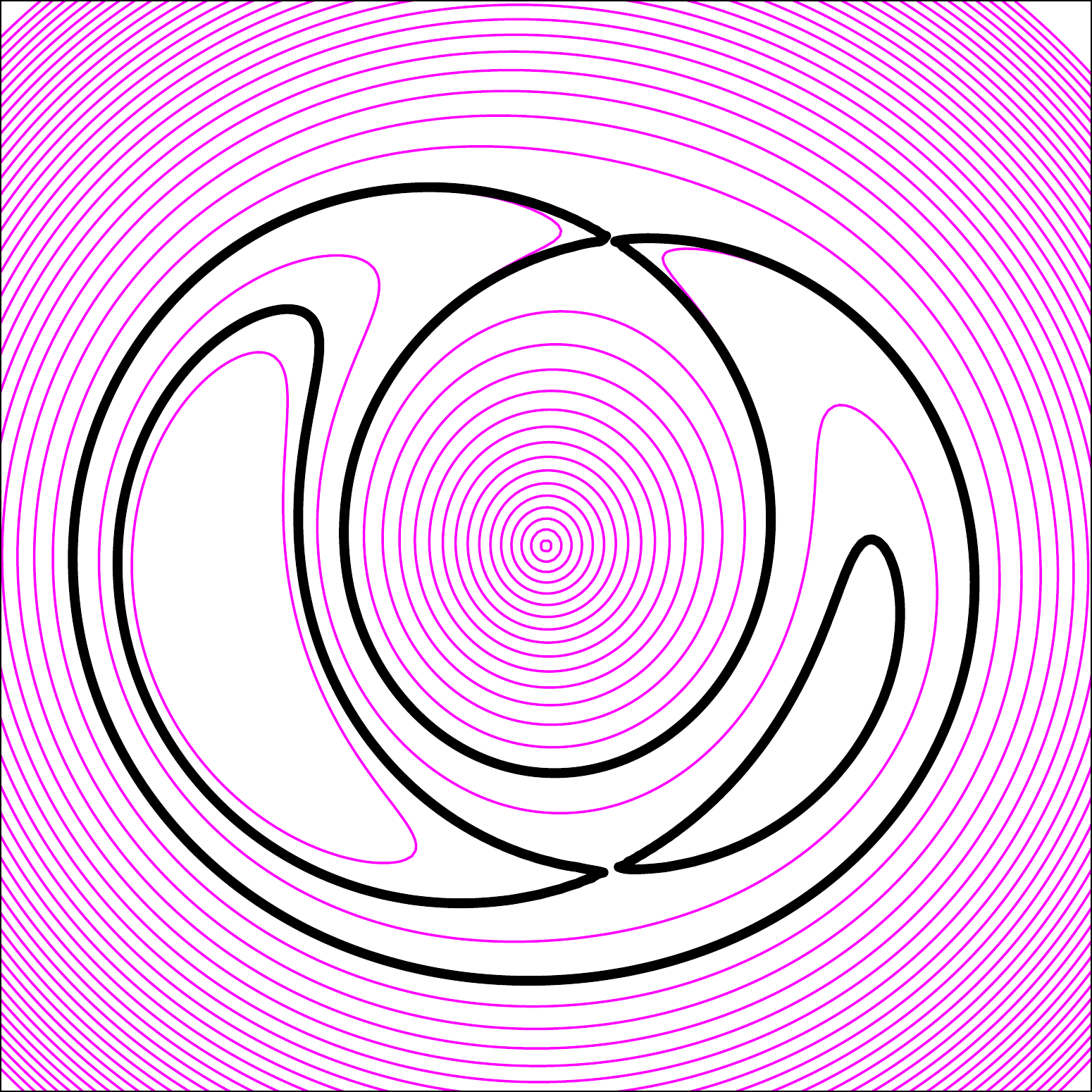}
  \includegraphics[width=\myplotswidth]{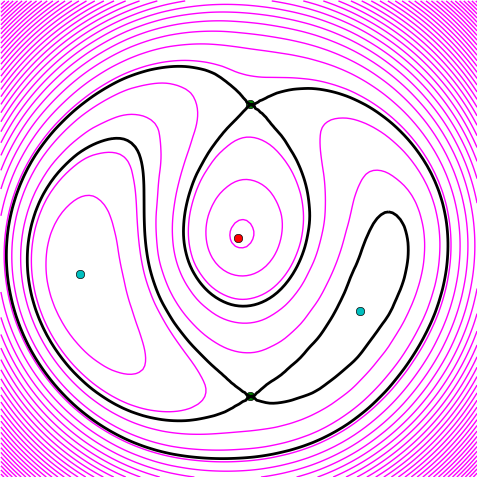} \\
  \includegraphics[width=\myplotswidth]{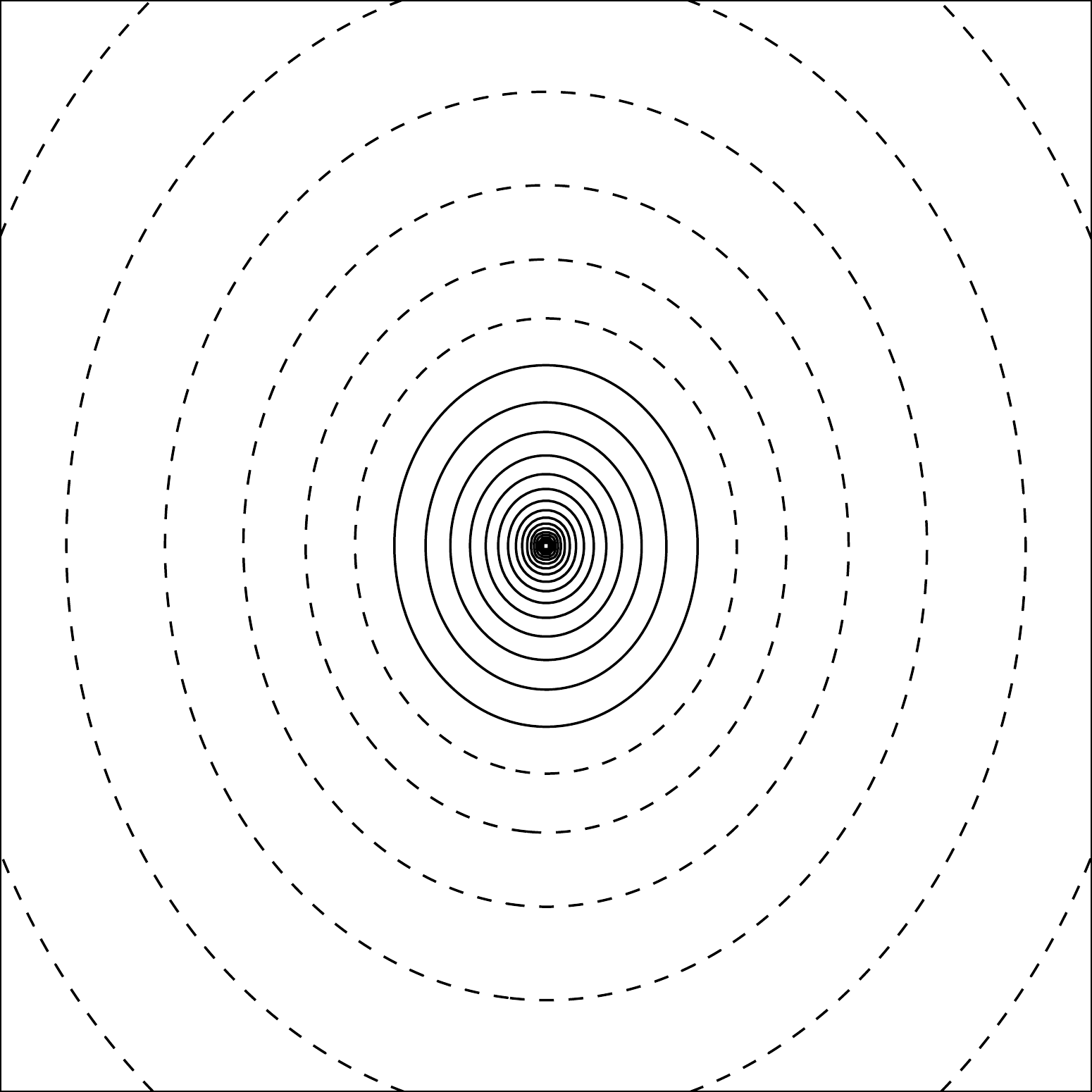}
  \includegraphics[width=\myplotswidth]{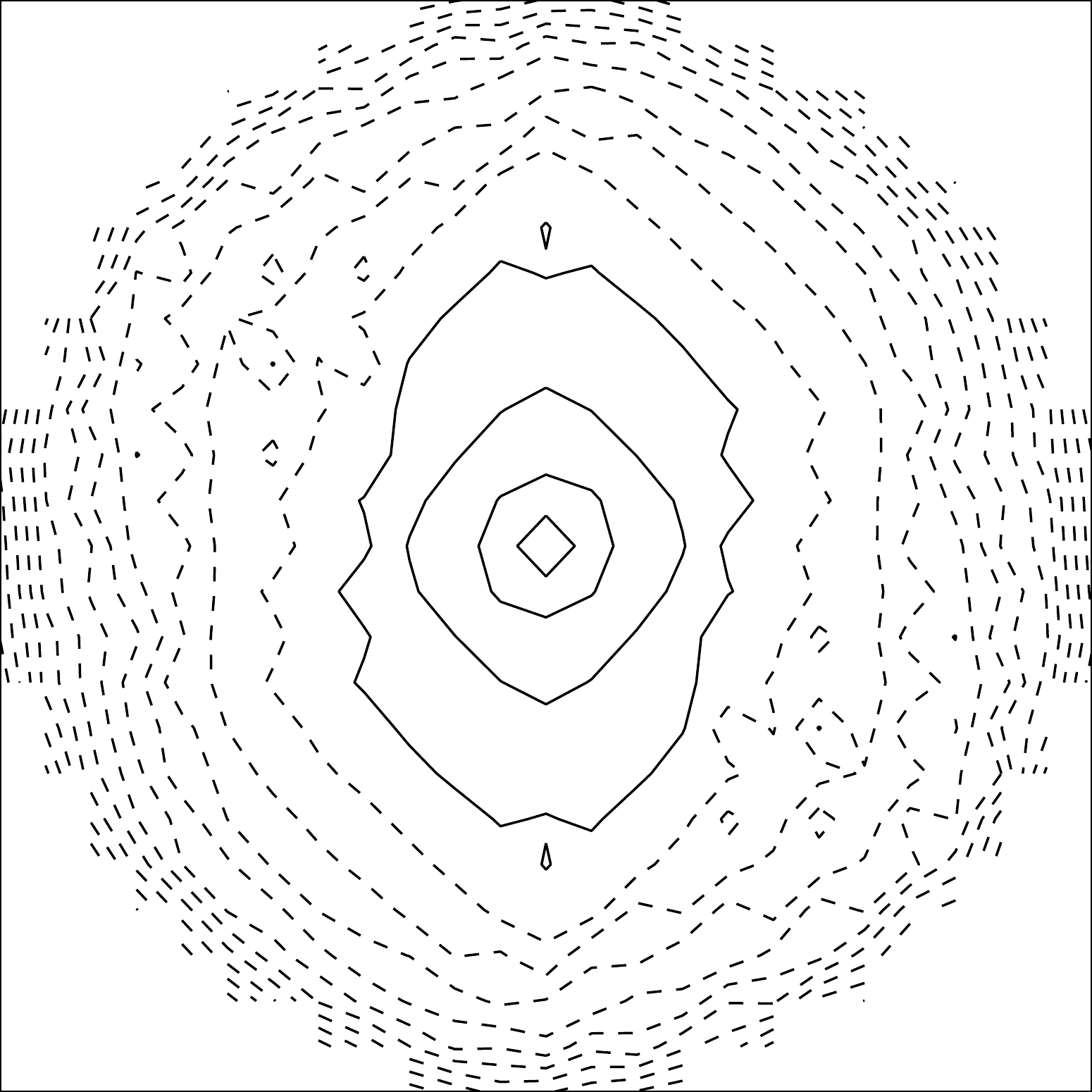}

  \caption[result 7022 (ASW0000h2m)]{The same system as in Figure
    \ref{fig:7025}, this time with image parities correctly
    identified. (See \secref{example_models} for details.)}
  \label{fig:7022}
\end{figure}

\FloatBarrier

The {\tt gravlens} program \citep{2001astro.ph..2340K} was used.
Formulas for the lenses appear in \cite{2001astro.ph..2341K}. The SIE
lenses follow equations (33--35) of that work, with core radius set to
zero.  The NFW lens is in equations (48) and (50), while shear is the
$\gamma$ term in equation (76).

The information in this section was not revealed to the main developer
of \spl (RK, who also chose the challenge set) or to the modellers
(EB, CC, CM, JO, PS and JW) while modelling was in progress.  That is,
the modellers had no advance knowledge of what kind of
parameterisation had been used to make the sims.  After the modelling
stage, AM released the details of the sims for post-modelling
analysis.  Results from the latter now follow.


\subsection{Some example models} \label{sec:example_models}

Of the 119 models proposed, we now discuss eight examples in some
detail.  Results from these are shown in Figures
\ref{fig:6941}--\ref{fig:7022}.  The first four of these show the most
common image morphologies, the other four explain some problem cases.

Each of Figures \ref{fig:6941}--\ref{fig:7022} figures has the
following layout.
$$ \begin{matrix}
\hbox{marked-up CFHTLS image} \qquad &\hbox{model synthetic image} \\
t(x,y)                               &\hbox{model } t(x,y) \\
\kappa(x,y)                          &\hbox{model } \kappa(x,y)
\end{matrix} $$

The model synthetic image presented in this paper is not the original one,
but an interpolated version generated by an updated version of \spl. The image
presented during the original experiment was of lower resolution. The two plots in
the middle showing $t(x,y)$ have uniform, but arbitrary spaced contour lines.
The $\kappa(x,y)$ plots in the bottom row show solid lines for $\kappa>1$
and dashed lines for $\kappa<1$. The spacing is logarithmic, with 10 contours
for every decade --- that means contour spacing is a quarter magnitude in
optical terms.

Let us now consider these cases in turn.

\begin{itemize}

\item Figure \ref{fig:6941} shows the simplest case, with two clear
  images produced by a nearly-circular lens.  The center of the
  lensing galaxy is a maximum, the image nearer to the galaxy is a
  saddle point, and the image further away is a minimum.  All these
  were correctly identified.  As note above, in
  \secref{SpaghettiLens}, the precise shape of the loops in the
  spaghetti diagrams is unimportant, only the implied image locations,
  parities and time-ordering matters. $\kappa(x,y)$ shows, that
  the model has a more shallow mass distribution than the simulation.
  This is a persistent issue throughout all models and is discussed in
  \secref{tests.t2}.

\item Figure \ref{fig:6990} shows an example of an arc that has split
  into three images.  This kind of configuration, with a counter-image
  close to the lensing galaxy and a more distant arc/triplet on the
  other side, generically arises from an elongated mass distribution
  when the source is displaced along the elongated direction.  The
  spaghetti diagram in this case has another markup element, a grey
  point and circle overlaid on a probable secondary lensing galaxy.
  This is an instruction to \spl to allow a point mass at that
  location, distinct from the main mass map.

\item Figure \ref{fig:6919} shows another example of an arc plus
  counter-image, but (in contrast to \figref{6990}) the arc is closer
  to the lens than the counter-image. This configuration arises if the
  source displacement is perpendicular to the long axis of the lensing
  mass.  Comparing the two panels in the middle row, we see that the
  modeller interpreted arc as consisting of three images, whereas the
  sim shows a single saddle point associated with the arc.  But the
  identification is not really erroneous --- we just need to take into
  account that the source is extended.  In fact, in the sim the
  brightest part of the source is only doubly imaged, but the source
  extends into a region that produces four images.  In the $t(x,y)$ of
  the sim, the hairpin-bending contours are typical of double on the
  verge of splitting into a quad.

\item Figure \ref{fig:6915} shows another quad.  This kind of
  configuration arises when the mass is elongated and the source is
  displaced at an angle to the elongation.  The minima and saddle
  points are correctly identified, and the orientation of the
  ellipticity of the mass distribution is correctly reproduced.

\item Figure \ref{fig:6975} shows a lens with substructure in the form
  of a smaller secondary galaxy.  The galaxies in such group or
  cluster sims were based on galaxies visible in the images, but the
  modelers were not told in advance whether this was the case.  The
  minimum and saddle point are correctly identified.  The mass
  distribution misses the substructure, but overall appears
  reasonable.

\item Figure \ref{fig:6937} shows a sim with substructre, like
  Figure \ref{fig:6937}. In comparison to the above, the resulting mass
  model is poor.

\item Figure \ref{fig:7025} shows a quad.  In this one, the
  identification of the minima and saddle points was incorrect, and
  mass distribution comes out elongated East-West instead of
  North-South.  The mass distribution also appears somewhat jagged and
  the saddle-point contours are not as clean as in the previous
  examples; these are often indicators of a problem with the model.
  The enclosed mass is, however, none the worse --- the reason is
  probably that in a relatively symmetrical image configuration, the
  Einstein radius is quite well constrained by the images in a fairly
  model-independent way.

\item Figure \ref{fig:7022} shows another model of the same system, the
  only one done by an expert in this sample. The image parities are
  correct. The elongation has the right orientation, but is too shallow.

\end{itemize}


\subsection{Test of image identification} \label{sec:tests.t1}

The first post-modelling test was a qualitative comparison of the
original arrival-time surfaces and the input spaghetti diagrams given
by the modeller.  This tested first, for correct identification and
location of the lensed images, and second, for the correct parities and
ordering of the lensed images in respect of the arrival time.

While we expected the identification of lensed images to be trivial,
given the generally clean appearance of the sims in the test, we
expected the parities and time-ordering to be more difficult.  While
the \spl tutorials had provided general guidelines, to be consistently
correct with the time-ordering, a modeller needs to develop some
intuition for arrival time surfaces.  This is an area where experience
and tutorials training could improve results at a later stage, and
correspondingly, feedback on the difficulties modellers encounter can
help improve the tutorial materials.

\tabref{stats} presents a summary of the test.  The evaluation was
done manually, comparing the input to \spl with the actual
arrival-time surface of the sim.  This amounts to comparing the
middle-left and middle-right panels in each of
Figures~\ref{fig:6941}--\ref{fig:7022}, and similarly for the other 111
models.

\begin{table}\centering\begin{tabular}{lrr}\hline
    Total                                                     & 119        & 100\%        \\
\hline
    errors in image locations          &     9         &    8\%            \\
    errors in image parities or time ordering     &    49     &    41\%     \\
\hline
    inaccurate image placement over an arc     &    21    &    18\%     \\
    identified two images of four                         &     5    &     4\% \\
    identified two nearby images as one & 3  & 3\%     \\
    missed faint images &   1   &  1\%     \\
    proposed too many images &  1  &  1\%    \\
    modelled a three-image arc as one image  &  4  &   3\%   \\
    modelled one image as a three-image arc  &  5  &   4\%   \\
    swapped minimum and saddle in double       &  2  &    2\%   \\
    swapped minima and saddles in quad            & 38 & 32\%   \\
    swapped early and late saddles in a quad &   7 &   6\%    \\
\hline
\end{tabular}
\caption{Table of image-identification errors and the number of models
  containing each.  A model can contain more than one type of error.}
\label{tab:stats}
\end{table}

The images of the system were considered to have been located
correctly, if all the images were identified and were approximated
within about 5\% of \spl frame used to draw the spaghetti diagram.
That frame size is adjustable by the user, but in practice it is
somewhat larger than the spaghetti diagram.  Such image-placement
errors were found in only 9 models.  That does not include inaccurate
image placement over an arc, which was considered a separate category
of error.

In addition to simple image-placement errors, ten types of errors were
recognised and are listed in \tabref{stats}.  Most of the problems
were due to unclear arc-like structures.  Critical errors like the
failure to identify all five images in a five images system, or to
include too many images, were rare.

The assignment of the parity of the images was a more difficult task,
and was successful in only about 60\% of the cases.  The most common
error was swapping of minima and saddle-points in a quad;
\figref{7025} shows an example.  Another, less common, error was
flipping the spaghetti diagram, thus swapping the time-ordering of the
two saddle points.

Incorrect image parities and time orderings tended to produce
poorer-looking models, such as the checker board patterns in the mass
map in \figref{7025}.  Interestingly, however, the enclosed-mass
profiles were quite robust.  We will consider this aspect in the next
section.

\FloatBarrier

\begin{figure}
  \centering
  \subfigure{\includegraphics[width=0.8\linewidth]
             {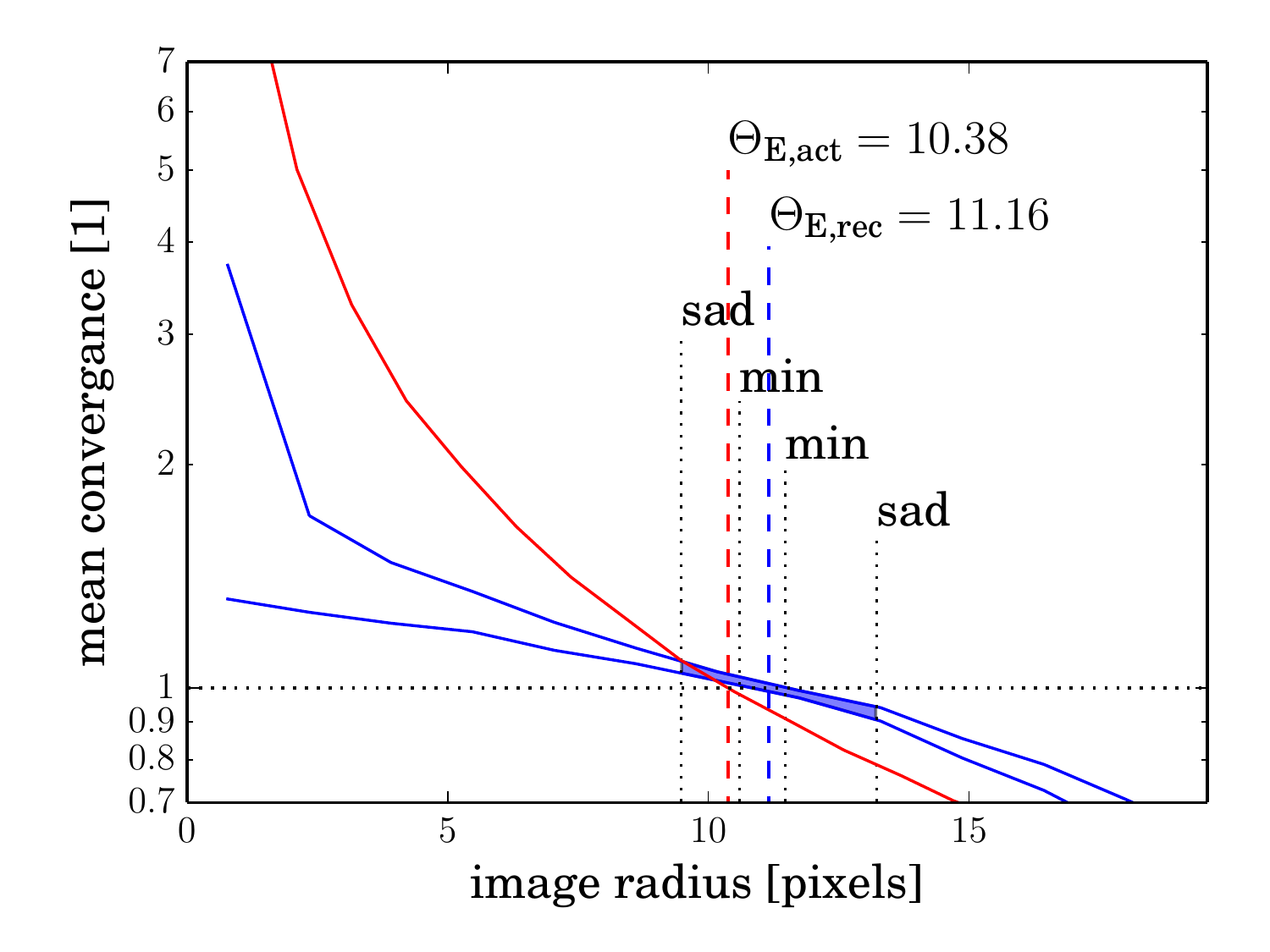}}
  \subfigure{\includegraphics[width=0.8\linewidth]
             {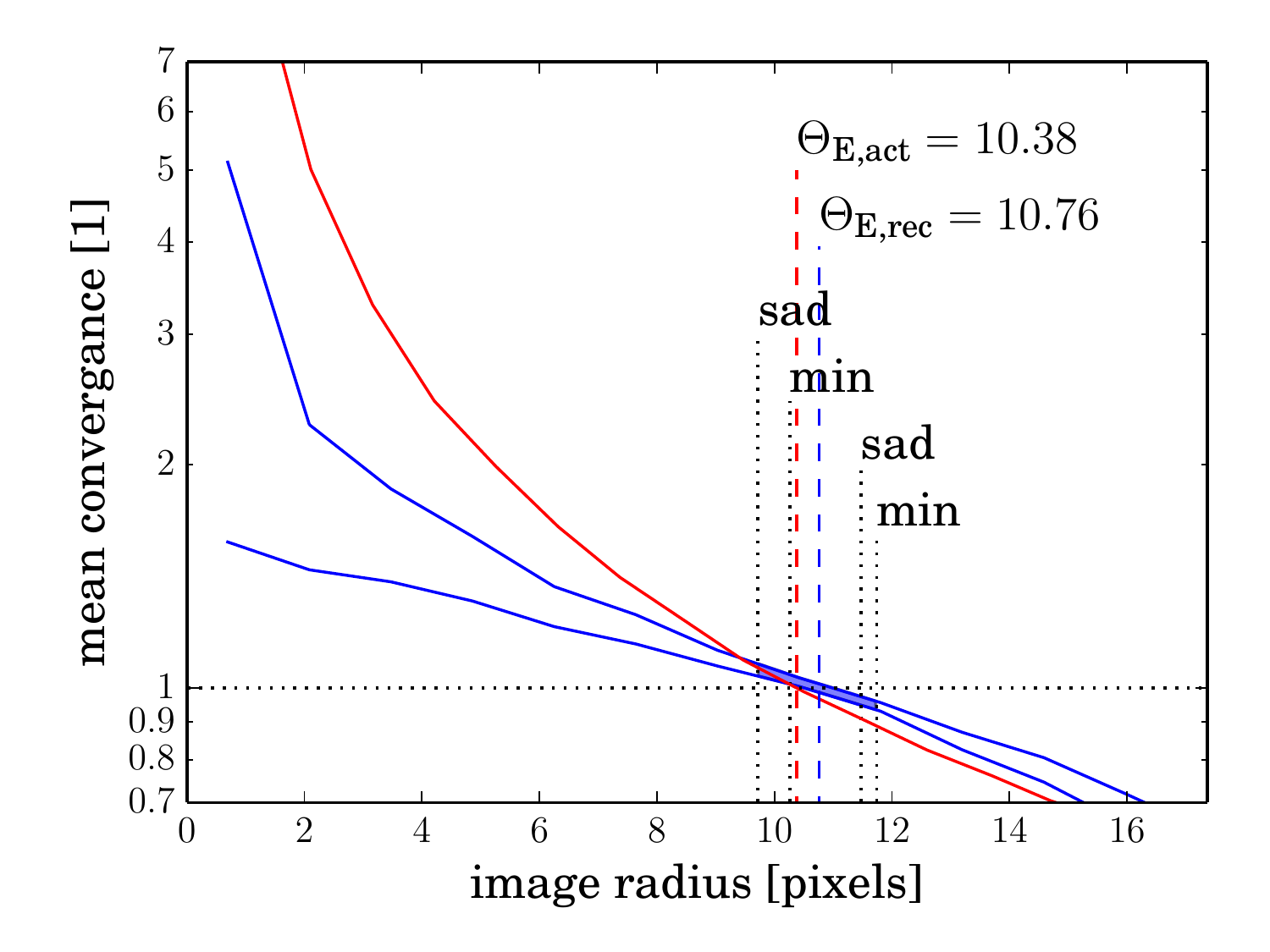}}
  \subfigure{\includegraphics[width=0.8\linewidth]
             {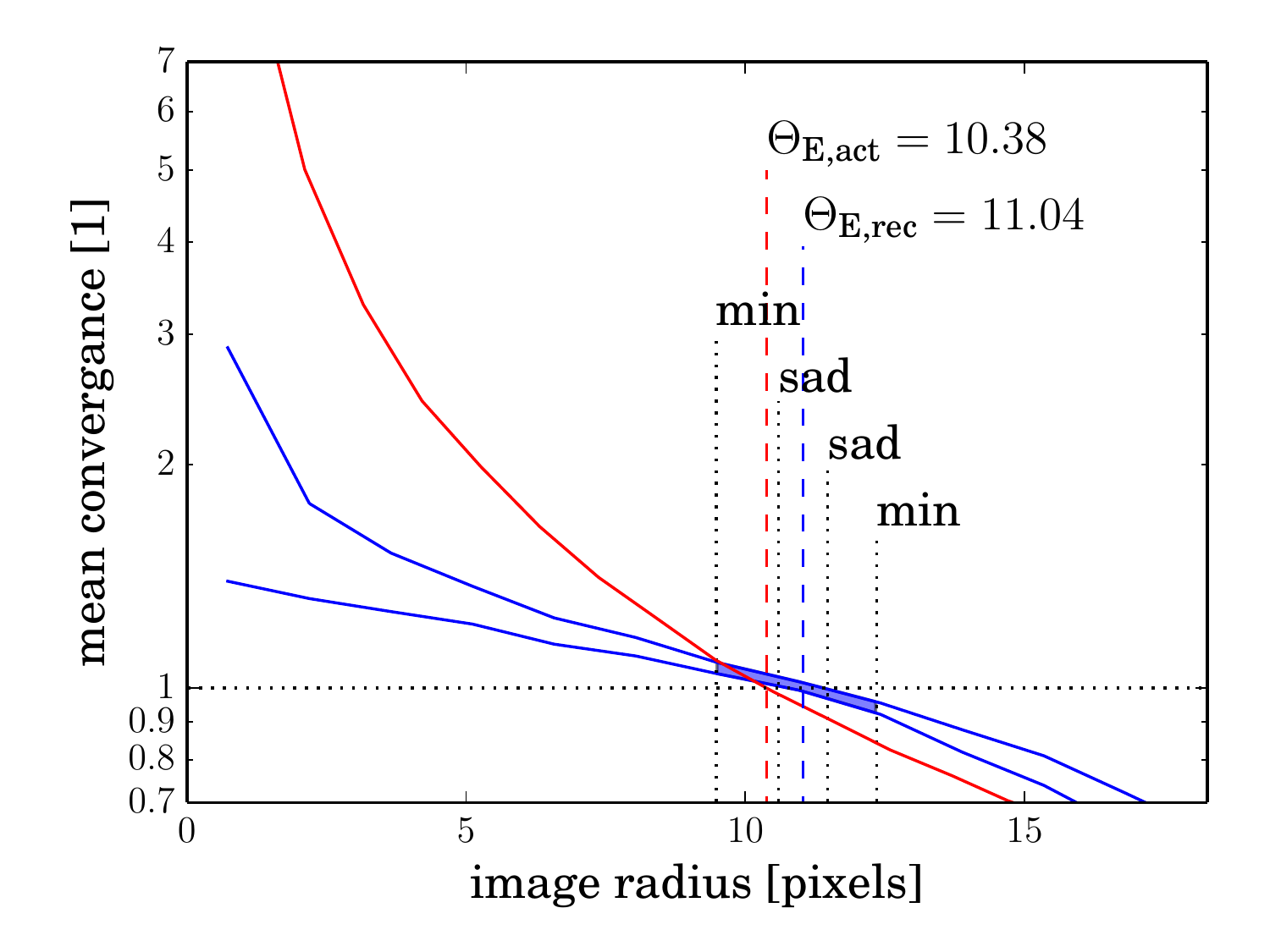}} \\
  \caption{Mean $\kappa$ inside a circle around the lens centre, as a
    function of the radius of the circle.  (See \secref{tests.t2} for
    details.) The upper panel corresponds to the model shown in
    \figref{7025}, in which the minima and saddle-points have been
    incorrectly swapped.  The middle corresponds to \figref{7022},
    where the image parities were correct.  The lower panel
    corresponds to another model, where the image parities were
    correct but the time-ordering was incorrect.}
\label{fig:kapenc_compare_faulty}
\end{figure}


\subsection{Test of mass-profile recovery} \label{sec:tests.t2}

The second test was to compare the mass distributions $\kappa(x,y)$ of
the sims and of the \spl models.  A visual comparison is presented for
the eight models in Figures~\ref{fig:6941}--\ref{fig:7022}, in the
lower-left versus lower-right panels.  We will summarize the mass
distributions drastically in a single number, the effective Einstein
radius.  Other measures for comparison of free-form lensing mass
distributions appear in \cite{2014arXiv1401.7990C}, but comparing
Einstein radii is already useful.

There is no standard way of defining the Einstein radius of a general
non-circular lens.  We adopt the simple definition
\begin{equation} \label{eq:effRE}
 \langle\kappa\rangle_{\Theta_{\rm E}} = 1,
\end{equation}
that is, the effective Einstein radius \ERf is such that the
mean $\kappa$ is unity inside a circle of radius \ERf centered at
the lens center.

\begin{figure*}
  \centering
    \includegraphics[width=0.90\linewidth]{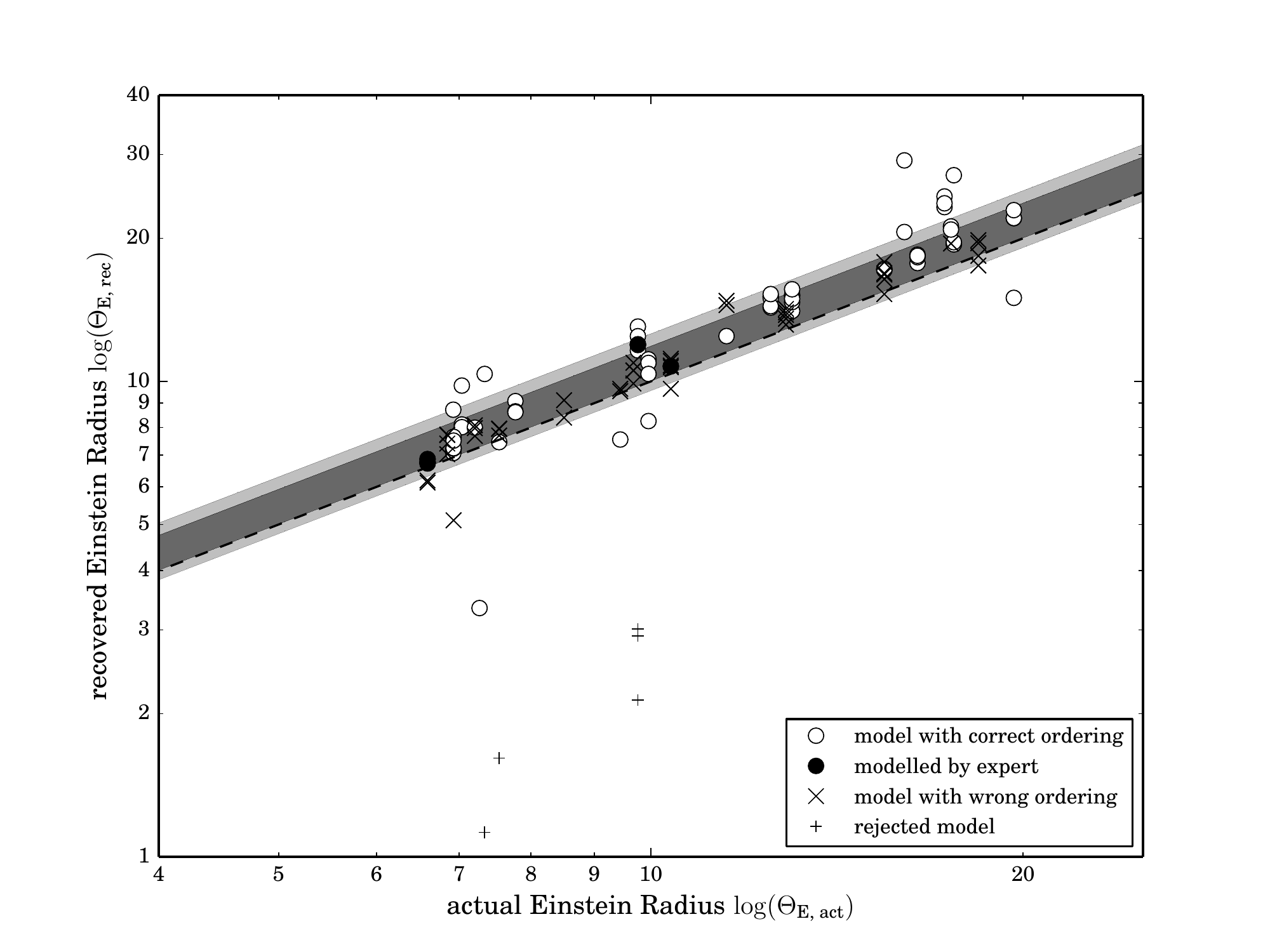}
  \caption{Model-recovered versus actual Einstein radii \ERf[,rec] and
    \ERf[,act].
    Plus signs indicate models flagged by the modeller as
    failures by commenting negatively about it in the forum. Light and dark grey
    bands show standard deviation of volunteers (15\%) and expert (10\%).}
  \label{fig:ER_per_sim}
\end{figure*}

To illustrate, \figref{kapenc_compare_faulty} compares the circularly
averaged mass profiles of three different models of one particular lens;
two of the models are shown in Figures~\ref{fig:7025} and
\ref{fig:7022}. Each panel in \figref{kapenc_compare_faulty} shows the
mean $\kappa$ within a circle of given radius.  The red curve is the
correct profile for the sim.  The two blue curves are the minimal and
maximal mean enclosed $\kappa$ from the internal ensemble in \spl.
Radial locations of the images are marked, along with the image
parities.  The region between the blue curves is shaded between the
radii of the innermost and the outermost images: this is the
confidence region from the modelling.  The definition \eqref{eq:effRE}
for \ERf corresponds to crossing the dashed horizontal line at $1$:
the red curve crosses the dashed line at the actual Einstein radius
\ERf[,act]; the recovered Einstein radius \ERf[,rec] and its
uncertainty are given by the blue curves crossing the dashed line.  We
see that in all three panels, the blue curves are shallower than the
red curve and \ERf[,rec] is more than \ERf[,act], by more than the
model uncertainties.  Now, steeper mass profiles tend to give wider
image separations --- recall that the image separation for a circular
isothermal lens is 2\ERf, whereas for a point mass it is more
\citep[see, e.g.,][]{2002LNP...608....1C} --- so \ERf[,rec] being too
high is really a consequence of the GLASS models being too shallow for
the sims.

\Figref{ER_per_sim} shows that \ERf[,rec] of the models tend to be too
high.  However, this is entirely due to the GLASS model density
profiles being too shallow, as illustrated above. We can separate out
the performance of the \spl interface and its users by comparing their
results with the Einstein radii of \spl models made by an expert
(PS). Discounting the models which were flagged by the volunteers as
poor, the mean Einstein radius overestimate was 10\%, with a 15\%
standard deviation (shown by the light grey band in
\Figref{ER_per_sim}).  The expert models show a similar bias, with
standard deviation 10\% (the dark grey band in \Figref{ER_per_sim}).
One source of this systematic error is that it is difficult to centre
the lens accurately: an offset leads to a flatter mass profile for the
model compared to the simulation.


\section{Outlook} \label{sec:todo}

This work has developed the concept of saddle-point contours in the
travel time of virtual photons, originally introduced by
\cite{1986ApJ...310..568B} for understanding image structure in strong
gravitational lenses, into a technique for mass-mapping lenses.
Despite being highly abstract, saddle-point contours look like
schematic arcs, and hence lend themselves to an intuitive markup tool
for lenses or lens candidates, which we call a spaghetti diagram.  At
the same time, saddle-point contours encode information about possible
mass distributions, which can be translated into input for an existing
lens-modelling engine \citep[GLASS, by][]{2014arXiv1401.7990C}.

\spl is an implementation of these ideas, enabling experienced but
non-professional lens enthusiasts to model newly-discovered lens
candidates from the \sw citizen-science platform.  The tests in this
paper indicate that such modelling would be both feasible and
scientifically interesting: given a suitable modelling tool, and
appropriate guidance, a small team of non-professional volunteers was
able to model a sample of 29 test lenses, and measure their Einstein
radii with comparable accuracy to a professional expert.

There is, however, plenty of room for improvement:
\begin{enumerate}
\item \spl tends to overestimate the Einstein radius (evident from
  \Figref{ER_per_sim}), and the model density profiles tend to be
  shallower than the SIE model used for generating the sims.  The
  likely explanation is that while the sims are steeply peaked at the
  centre, the pixellated mass model fixes a comparatively large area
  near the central at constant density. Allowing smaller pixels in the
  centre region, thus enabling a steeper centre \citep[similar to the
    ``high resolution'' feature implemented in][]{2014arXiv1401.7990C}
  may remove this bias.  The use of simply parameterised,
  appropriately steeply-profiled models would also avoid the problem.
\item Currently, \spl does not attempt to model the source shape; the
  user identifies the brightest points on the image, and these are
  taken as images of a point-like source, whose positions must be
  reproduced exactly. For generating a synthetic image, a conical
  source profile is assumed. Fitting for the source profile to
  optimize resemblance to the observed lensed image after the lens
  model has been generated, is algorithmically straightforward
  \citep[cf.][]{2003ApJ...590..673W,2006MNRAS.371..983S} and planned
  to be implemented.  This would alleviate another problem with \spl,
  which is that there is as yet no quantitative figure of merit for
  any given model: assessment of each model is a judgment call based
  on the synthetic image, and on whether the mass distribution and the
  arrival-time surface show suspicious features.  Another possibility
  would be use the \spl models as a feeder to a different
  lens-modelling program that already implements source-profile
  fitting.
\item Another limitation so far in \spl is that the lens is assumed to
  be dominated by one galaxy, which puts most galaxy-group lenses
  beyond the reach of the modeler. Since complicated group lenses are
  some of the most interesting candidates present, removing this
  limitation is most desirable.  From the users' point of view, it
  would mean that spaghetti contours with more than one maximum can be
  allowed.  For examples, see Fig. 5c in \citep{2001ApJ...557..594R}
  and Fig. 4b in \cite{2003ApJ...590...39K}.
\item At present, a single false-color composite is used as the data.
  An option could be added to use all available filters, individually
  or in combination, at the user wishes.
\item As mentioned above, the option of revising an already-archived
  model is already available.  Desired now are tools for comparing
  different models of a given system, both visually and through
  different statistical measures.  As evidenced by a current
  collaborative modelling effort, a particularly interesting candidate
  can lead to an extended discussion and dozens of models, that
  in some way sample the high likelihood region of model parameter
  space.
\item Better tutorial materials are also needed, and this would
  address some of the problem areas found in the modelling challenge.
  For example, we saw in \secref{tests.t1} that volunteers are most
  prone to making errors in two situations: when in identifying an arc-
  like structure while placing the points, and in identifying the
  correct ordering of the points in nearly-symmetric configurations.
  Better and more detailed introductory materials would also allow
  the community of modellers to grow faster and without individual
  instructions by experts or experienced volunteers.
\end{enumerate}

The \spl program was developed by Küng, with design suggestions from
Coles, Cornen and Saha, and feedback from all co-authors.  The
simulations were created by A.~More, in consultation with Marshall,
S.~More and Verma.  Modelling was done by Baeten, Cornen, Macmillan,
Odermatt, Saha and Wilcox, with post-modelling analysis by Küng and
Saha.  All authors participated in writing and editing the manuscript.


\section*{Acknowledgments}

We thank the Swiss Society for Astrophysics and Astronomy and the
Swiss Academy of Sciences. The work of AM and SM was supported by
World Premier International Research Center Initiative (WPI
Initiative), MEXT, Japan. AV is supported by a research fellowship
from the Leverhulme Trust. RK is supported by the Swiss National
Science Foundation. This work was supported in part by the U.S.
Department of Energy under contract number DE-AC02-76SF00515.


\appendix

\section{Relation to standard lensing formalism}\label{more-theory}

The description of the arrival-time surface in
Sections~\ref{sec:Fermat} and \ref{sec:arriv} omitted some details for
the sake of a more intuitive explanation.  The convergence $\kappa$
and the geometric time delay $t_{\rm geom}$ were left as
proportionalities (equations \ref{eq:kappa} and \ref{eq:tgeom}), and
the gravitational time delay $t_{\rm grav}$ was given in an implicit
form (equation \ref{eq:tgrav}).  Here we fill in the details.

The original formulation of the arrival-time surface appears in
equations (2.1) to (2.6) of \cite{1986ApJ...310..568B}.  Their
equations can be rearranged as follows.
\begin{equation} \label{cfBN86}
\begin{aligned}
t_{\rm geom} &= \frac{(1+z_L)}{2c} \frac{d_S}{d_L d_{LS}}
\left[ (x-x_s)^2 + (y-y_s)^2 \right] \\
\nabla^2 t_{\rm grav} &= -(1+z_L)\frac{8\pi G}{c^3} \, \Sigma(x,y) \\
\kappa(x,y) &= \frac{4\pi G}{c^2} \frac{d_L d_{LS}}{d_S}
               \times \Sigma(x,y)
\end{aligned}
\end{equation}
The symbols $d_L,d_S$ and $d_{LS}$ are angular-diameter distances,
respectively from observer to lens, observer to source, and lens to
source. We have replaced angular positions on the sky with positions
on the lens plane as
\begin{equation}
(x,y) = d_L (\theta_x,\theta_y) \,.
\end{equation}
In the concordance cosmology
\begin{equation}
d_{LS} = \frac{c}{H_0}\,\frac1{1+z_S} \,
\int_{z_L}^{z_S} \!\! \frac{dz}{\sqrt{\Omega_m(1+z)^3 + \Omega_\Lambda}}
\end{equation}
and similarly $d_L$ and $d_S$.

The first line of equation \eqref{cfBN86} is $t_{\rm geom}$ from
equation \eqref{eq:tgeom} with the proportionality filled in, and with
the source offset at $(x_s,y_s)$ rather than at the origin.

The last line of equation \eqref{cfBN86} fills in the proportionality
factor in equation \eqref{eq:kappa} for the convergence (or
dimensionless surface density) $\kappa$.

The middle line of equation \eqref{cfBN86} is a Poisson equation for
the gravitational time delay, and is equivalent to the implicit
expression \eqref{eq:tgrav}.  One way to verify the equivalence is to
consider the small circle in equation \eqref{eq:tgrav} as a region
where $\Sigma$ is constant, and approximate $\tgrav$ by its Taylor
expansion to $O(x^2,y^2)$.  Substituting in equations \eqref{cfBN86}
gives the Taylor coefficients in terms of $\Sigma$, and result
satisfies the expression \eqref{eq:tgrav}.  Alternatively, we can
proceed with a discrete form of the Poission equation from
\eqref{cfBN86}.  Discretising on a grid with spacing $\Delta$, we have
\begin{equation} \label{eq:relax}
\begin{aligned}
\tgrav(x,y) &=
\frac14 \Big[ \tgrav(x+\Delta,y) + \tgrav(x-\Delta,y) + {}\\
&\kern24pt    \tgrav(x,y+\Delta) + \tgrav(x,y-\Delta) \Big] \\
&+ (1+z_L)\frac{2G}{c^3} \, \pi\Delta^2 \, \Sigma(x,y) \,.
\end{aligned}
\end{equation}
This is recognisable as a formula for solving the two-dimensional
Poisson equation from \eqref{cfBN86} by relaxation.  Let us now
replace the average over four neighbouring points by the circular
average $\left\langle \tgrav(x_\subcirc,y_\subcirc) \right\rangle$ and
replace $\pi\Delta^2\,\Sigma(x,y)$ by the enclosed mass
$M(x_\subbullet,y_\subbullet)$.  These replacements are valid in the
limit of a small grid.  The result is the implicit
equation \eqref{eq:tgrav}.


\bibliographystyle{mn2e}
\bibliography{ms}


\end{document}